\documentclass[prd,amsmath,amssymb,longbibliography,12pt,nofootinbib]{revtex4-2}
\usepackage[normalem]{ulem}
\usepackage{soul}
\usepackage[utf8]{inputenc}
\usepackage{lineno,hyperref}
\usepackage{amsfonts,amssymb,amsmath}
\usepackage{epsfig}
\usepackage{mathrsfs}
\usepackage{amssymb}
\usepackage{graphicx}
\usepackage{amsmath}

\usepackage{xcolor}
\usepackage{color}
\usepackage[normalem]{ulem}
\usepackage{soul}
\usepackage{appendix}

\usepackage{xcolor}
\begin{document}

\title{Circuit complexity of coupled four-mode squeezed states in two-field inflation}
\author{Tao Li$^{1}$}
\email{2022700450@stu.jsu.edu.cn}
\author{Hai-Bing Fu$^{1}$}
\author{Lei-Hua Liu$^{2}$}

\affiliation{$^{1}$Department of Physics, Guizhou Minzu University, Guiyang 550025, China}
\affiliation{$^{2}$Department of Physics, College of Physics, Mechanical and Electrical Engineering, Jishou University, Jishou 416000, China}

\begin{abstract}

We study the circuit complexity of coupled curvature and entropy perturbations in two-field inflation. We start from the quadratic action and derive the full Hamiltonian. The Hamiltonian includes all diagonal and off-diagonal quadratic terms. We then obtain three coupled equations for the pair amplitudes of the four-mode squeezed state. From these amplitudes, we calculate the two principal squeezing strengths. We use them to determine the Nielsen circuit complexity. For an equal-weight Gaussian circuit, the complexity takes the form \(\mathcal{C}_{2,k}=\sqrt{2(r_{1k}^{2}+r_{2k}^{2})}\). It depends on the squeezing strengths but not explicitly on the squeezing angles. Our numerical results show that the complexity grows after horizon crossing in both coupled and uncoupled models. Interfield coupling changes the pair amplitudes and the four-mode squeezed state. However, its effect on the circuit complexity is small for the parameters considered. This result shows that different four-mode pair-amplitude matrices can yield similar values of the chosen equal-weight circuit complexity. The present calculation provides a reference for studying Gaussian circuit complexity in more general two-field inflationary models.

\end{abstract}

\maketitle{}
\section{Introduction}
\label{sec:introduction}

Inflation provides a mechanism for generating primordial perturbations from quantum fluctuations, but its success does not fix the number of scalar degrees of freedom active during this stage. Additional fields arise in assisted and $N$-flation constructions and in more general multifield models \cite{Liddle:1998jc,Dimopoulos:2005ac,Dias:2017gva}. Studies of random potentials, nonminimal couplings, and numerical transport illustrate the range of background trajectories and perturbation histories in such systems \cite{Tye:2008ef,Easther:2013rva,Dias:2016rjq,Dias:2015rca,Seery:2012vj}. In the adiabatic--entropy decomposition, a bend of the trajectory can transfer entropy fluctuations into the curvature perturbation and generate correlations between them \cite{Gordon:2000hv,Wands:2002cj,Langlois:2008mn}. This transfer has been studied in double inflation, multifield slow-roll theory, and superhorizon evolution \cite{Langlois:1999dw,GrootNibbelink:2001qt,Sasaki:1998ug,Wands:2002bn,Wands:2007bd}. Both the trajectory and the field-space metric enter the dynamics of coupled perturbations \cite{Gong:2011uw,Peterson:2011yt,Kaiser:2013sna}. Models with three or more fields can have several independent entropy directions. Two fields provide the minimal setting in which curvature--entropy transfer and interfield quantum correlations can be studied while retaining both perturbations.

A turning trajectory can make heavy-field effects relevant even when the background evolution appears approximately single-field. Depending on the masses and adiabaticity, integrating out heavy fluctuations can modify the effective sound speed and produce features in lower-energy observables \cite{Achucarro:2010jv,Achucarro:2012sm,Chen:2009zp,Achucarro:2012yr,Cespedes:2012hu,Gao:2012uq,Shiu:2011qw,Tolley:2009fg}. Field-space curvature can also affect the stability of the trajectory \cite{Renaux-Petel:2015mga}. Noncanonical kinetic terms, DBI interactions, and nonminimal couplings further broaden the possible perturbation dynamics \cite{Langlois:2008mn,Langlois:2008qf,Kaiser:2012ak}. Curvature--entropy conversion can affect power spectra and adiabatic--isocurvature correlations, while nonlinear evolution can contribute to higher-order statistics \cite{Lyth:2005fi,Vernizzi:2006ve,Byrnes:2008wi,Byrnes:2010em,Peterson:2010mv}. Bispectra and trispectra in multifield models have been studied using perturbative and transport methods \cite{Gao:2008dt,Arroja:2008yy,Anderson:2012em,Elliston:2012wm,Dias:2016slx}. CMB measurements constrain surviving isocurvature contributions \cite{Planck:2018jri}. These results motivate retaining the interfield terms in the quadratic action under consideration. Circuit complexity characterizes a state-preparation problem. Relating it to an observed spectrum or higher-order correlator requires a separate calculation.

Linear evolution of inflationary quantum fluctuations produces squeezed Gaussian states \cite{Albrecht:1992kf,Grishchuk:1990bj}. For one uncoupled perturbation, the opposite-momentum modes $(k,-k)$ form a two-mode squeezed state. With two coupled perturbations, the same momentum pair contains curvature and entropy modes at both momenta. Same-field and cross-field pair creation then belong to one four-mode squeezed state. Entanglement and quantum discord provide other descriptions of cosmological quantum correlations \cite{Bolis:2018jmo,Martin:2015qta}. A four-mode construction includes both squeezing and interfield transfer in coupled two-field systems \cite{Colas:2021llj}. Canonical phase-space methods provide a broader framework for coupled Gaussian modes \cite{Nilles:2001fg,Hackl:2020ken}. Principal squeezing channels can be identified from the joint state at a given time, but they need not coincide with the original curvature and entropy fields or dynamically decouple the Hamiltonian.

Circuit complexity measures the minimum cost of preparing a target state from a specified reference using a chosen set of gates \cite{Nielsen:2005mkt,Dowling:2006tnk}. Wavefunction and covariance-matrix methods implement this geometric construction for Gaussian quantum fields \cite{Jefferson:2017sdb,Chapman:2018hou,Chapman:2017rqy}. Studies of fermions, interactions, quantum quenches, and different complexity prescriptions show why the reference, gates, and cost function must be specified \cite{Hackl:2018ptj,Bhattacharyya:2018bbv,Alves:2018qfv,Camargo:2018eof,Ali:2018fcz}. In the equal-weight Gaussian geometry used here, the minimum distance depends on the principal squeezing strengths and is independent of squeezing angles when those strengths are held fixed. Cosmological circuit complexity has been studied for squeezed perturbations, different expanding backgrounds, non-trivial sound speed, modified dispersion relations, and $k$-essence dynamics \cite{Bhattacharyya:2020rpy,Bhattacharyya:2020kgu,Liu:2021nzx,Li:2023ekd,Li:2021kfq}. Krylov and spread complexity have been investigated in related field-theory and cosmological settings \cite{Li:2024kfm,Li:2024iji,Li:2024ljz,Caputa:2021sib,Balasubramanian:2022tpr,Adhikari:2022oxr,Adhikari:2022whf,Zhai:2024odw}. These quantities describe spreading under a chosen evolution and do not determine a geometric state-preparation distance. The modified-gravity calculation \cite{Li:2026jxx} and the two-field inflationary Krylov analysis \cite{Li:2026cry} provide related points of comparison. 

We apply the established equal-weight Gaussian-state Nielsen distance \cite{Chapman:2018hou} to the coupled four-mode squeezed state of two-field inflation. Starting from the quadratic action for curvature and entropy perturbations, we determine every diagonal and off-diagonal entry of the pair-annihilation matrix $A_{ij,k}$ and the number-conserving matrix $B_{ij,k}$. The complex conjugate of $A_{ij,k}$ multiplies the pair-creation terms. Three coupled complex equations determine the symmetric pair-amplitude matrix $Z_k$, including $Z_{12,k}$. The full matrix determines the two principal squeezing strengths, their phase data, and the interfield mixing information. For a vacuum reference and unrestricted quadratic Gaussian gates, we calculate the equal-weight Nielsen $F_2$ complexity of one complete momentum pair. We also formulate a weighted cost and distinguish its bounds and trial-path length from an exact minimum. Finally, a matched uncoupled evolution shows how interfield coupling changes the four-mode squeezed state and its circuit complexity.

The paper is organized as follows. Sec. \ref{sec:complexity} will define the circuit complexity of the four-mode Gaussian state. Sec. \ref{sec:dynamics} will derive the coupled squeezing dynamics and compares the resulting complexity with its uncoupled counterpart. Sec. \ref{sec:conclusion} will present the conclusions. The appendices will provide the wave-function construction, the complete Hamiltonian coefficients, and details of the squeezing parameter derivation.

\section{Circuit complexity of the four-mode squeezed state}
\label{sec:complexity}

For each nonzero momentum pair $(k,- k)$, the curvature and entropy perturbations form four quantum modes. For the homogeneous quadratic system considered here, the pure state of one opposite-momentum pair can be represented as a four-mode squeezed state. A four-mode construction~\cite{Colas:2021llj} describes its evolution through interfield transfers and opposite-momentum pair creation. Acting on the initial vacuum, the state can be written in Bloch--Messiah form as
\begin{equation}
 \begin{aligned}
 |\Psi_{T,k}\rangle&=\hat R_{L,k}\hat S_{1,k}(r_{1k},\phi_{1k})
 \hat S_{2,k}(r_{2k},\phi_{2k})|0_k\rangle,\\
 \hat S_{i,k}(r,\phi)&=\exp\!\left[r\left(
 e^{-2i\phi}b_{i,+}b_{i,-}
 -e^{2i\phi}b_{i,+}^{\dagger}b_{i,-}^{\dagger}\right)\right].
 \end{aligned}
 \label{eq:stateBM}
\end{equation}
Here \(b_{i,\pm}\equiv b_{i,\pm{k}}\). The squeezing operator convention agrees with Ref.~\cite{Li:2026jxx}. Each independent channel contributes $(-e^{2i\phi_{ik}}\tanh r_{ik})^n$ to its $n$-pair amplitude. The remaining passive factor on the right acts on $|0_k\rangle$ only by a phase. The nonnegative $r_{1k}$ and $r_{2k}$ characterize the coupled four-mode squeezed state. In the presence of interfield mixing, they cannot be assigned to the individual curvature and entropy fields.

In either real cosine or sine sector, set $ X=(Q_1,Q_2,P_1,P_2)^T$ and $G_{AB}=\langle\{X_A,X_B\}\rangle/2$. The chosen vacuum has covariance $I_4/2$. Reality, homogeneity, and the parity-even action below imply identical dynamics in the two real sectors. Thus $\mathbb G_R=I_8/2$ and $\mathbb G_T=G_{T,k}\oplus G_{T,k}$. The symplectic decomposition of one sector yields
\begin{equation}
 S_k=O_{L,k}\text{diag} (e^{r_{1k}},e^{r_{2k}},e^{-r_{1k}},e^{-r_{2k}})O_{R,k},\qquad
 G_{T,k}=\frac12S_kS_k^T.
 \label{eq:BM}
\end{equation}
The matrices \(O_{L,k}\) and \(O_{R,k}\) are orthogonal and symplectic. The right factor cancels from \(S_kS_k^{T}\). We choose the vacuum as the reference state and allow Hermitian quadratic gates
\(\hat{\mathcal G}_I=\tfrac12\hat X^{T}H_I\hat X\), where each \(H_I\) is real and symmetric. These gates act on the eight real quadratures \(\hat X=(Q_{1,c},Q_{2,c},P_{1,c},P_{2,c},
Q_{1,s},Q_{2,s},P_{1,s},P_{2,s})^{T}\) \cite{Nielsen:2005mkt,Dowling:2006tnk}. The unitary circuit and its real symplectic representation obey
\begin{equation}
 \frac{d\hat U}{ds}=-i\sum_I Y^I(s)\widehat{\mathcal G}_I\widehat U(s),\quad
 \frac{d\mathbb S}{ds}=\left[\sum_IY^I(s)\mathbb J\mathsf H_I\right]\mathbb S(s),\quad
 \mathbb J=J_4\oplus J_4,\quad J_4=\begin{pmatrix}0&I_2\\-I_2&0\end{pmatrix}.
 \label{eq:Nielsencontrols}
\end{equation}
The controls $Y^I(s)$ select field-transfer, interfield squeezing, and passive rotation gates. With a Frobenius-orthonormal symplectic generator basis, $\tfrac12\text{Tr}_8[(\mathbb J\mathsf H_I)(\mathbb J\mathsf H_J)^T]=\delta_{IJ}$, Nielsen's quadratic cost with gate weights $w_I>0$ becomes
\begin{equation}
 \mathcal C_{\boldsymbol w,k}=\inf_{\{Y^I\}}\int_0^1 ds\,
 \left[\sum_Iw_I^2\bigl(Y^I(s)\bigr)^2\right]^{1/2},\qquad
 \mathbb S(0)=I_8,\quad \mathbb S(1)\mathbb S(1)^T=2\mathbb G_T.
 \label{eq:Nielsenweighted}
\end{equation}
The minimization is over circuits preparing the state. The right passive factors at the endpoint are unconstrained. The desired gate set and its weights must be specified before a numerical complexity can be assigned.

For equal weights $w_I=1$ and all quadratic Gaussian generators allowed, Eq.~\eqref{eq:Nielsenweighted} reduces to the right-invariant $F_2$ length of the eight-dimensional symplectic circuit~\cite{Chapman:2018hou}
\begin{equation}
 \mathcal C_{2,k}=\inf_{\mathbb S(s)}\int_0^1ds\,
 \left[\frac12\text{Tr}_8\!\left(\dot{\mathbb S}\mathbb S^{-1}
 (\dot{\mathbb S}\mathbb S^{-1})^T\right)\right]^{1/2},
 \quad \mathbb S(0)=I_8,\quad \mathbb S(1)\mathbb S(1)^T=2\mathbb G_T.
 \label{eq:cost}
\end{equation}
The circuit coordinate $s$ is independent of conformal time. Since $2\mathbb G_T$ is positive and symplectic, $\mathbb S_*(s)=\exp[\tfrac{s}{2}\log(2\mathbb G_T)]$ prepares the target state. The singular-value bound derived in Appendix~\ref{app:covariance} shows that this path has minimum length. The four eigenvalues $e^{2r_{1k}},e^{-2r_{1k}},e^{2r_{2k}},e^{-2r_{2k}}$ each occur twice in $2\mathbb G_T$. Therefore, the minimum complexity is
\begin{equation}
 \mathcal C_{2,k}=\frac{1}{2\sqrt2}
 \left\{\text{Tr}_8\left[\bigl(\log(2\mathbb G_T)\bigr)^2\right]\right\}^{1/2}
 =\sqrt{2\left(r_{1k}^2+r_{2k}^2\right)}.
 \label{eq:Crr}
\end{equation}
This is the complexity of one complete momentum pair under the equal-weight Nielsen cost. It can equivalently be written as $\mathcal C_{2,k}(r_{1k},r_{2k},\phi_{1k},\phi_{2k})=\sqrt{2(r_{1k}^2+r_{2k}^2)}$. At fixed squeezing strengths, the equal-weight covariance-matrix distance is independent of the angles, unlike the single-channel expression obtained with a different wave-function circuit prescription in Ref.~\cite{Li:2026jxx}. With $w_{\min}=\min_Iw_I$, and with the positive radial path of Eq.~\eqref{eq:cost} as a trial circuit, one has the rigorous bounds
\begin{equation}
 w_{\min}\sqrt{2(r_{1k}^2+r_{2k}^2)}
 \leq\mathcal C_{\boldsymbol w,k}
 \leq \left[\sum_Iw_I^2\bigl(Y^I_{\rm rad}\bigr)^2\right]^{1/2},\qquad
 \sum_IY^I_{\rm rad}\mathbb J\mathsf H_I=\tfrac12\log(2\mathbb G_T).
 \label{eq:weightedbounds}
\end{equation}
The upper bound is computable directly from the complete covariance in Appendix~\ref{app:covariance}. For dynamically uncoupled channels, assign weights \(w_D\) and \(w_X\) to the diagonal and off-diagonal generators in each channel's \((Q_i,P_i)\) plane, respectively. The resulting radial trial length depends explicitly on the squeezing angles
\begin{equation}
 \mathcal C_{\boldsymbol w,k}\leq L_{\rm rad,k}
 =\left\{2\sum_{i=1}^{2}r_{ik}^2
 \left[w_D^2\cos^2(2\phi_{ik})+w_X^2\sin^2(2\phi_{ik})\right]\right\}^{1/2}.
 \label{eq:weightedradial}
\end{equation}
Eq.~\eqref{eq:weightedradial} gives the exact length of the specified radial path for uncoupled channels and hence an upper bound on their weighted complexity. It does not establish that this path minimizes the cost when \(w_D\ne w_X\). For coupled channels, the corresponding radial upper bound must instead be evaluated from the full target covariance and a gate basis with specified interfield weights, as in Eq.~\ref{eq:weightedbounds}.

\section{Complete squeezing dynamics of coupled inflationary perturbations}
\label{sec:dynamics}

In units of \(M_{\rm Pl}=\hbar=1\), the quadratic action for the coupled two-field perturbations takes the form~\cite{GrootNibbelink:2001qt,Li:2026cry}
\begin{equation}
 S^{(2)}=\int d\tau\,d^3x\,a^2\Biggl[
 \epsilon\left(\xi'-\frac{\mathcal H\lambda}{\sqrt{2\epsilon}}\psi\right)^2
 -\epsilon(\nabla\xi)^2+\frac12\psi'^2-\frac12(\nabla\psi)^2
 -\frac12a^2\mu^2\psi^2\Biggr].
 \label{eq:action}
\end{equation}
Primes denote conformal time ($\tau$) derivatives, $\mathcal H=a'/a$ is the conformal Hubble rate, $H=\mathcal H/a$ is the physical Hubble rate, and $\lambda=2\Omega_{\rm turn}/H$. The background functions $a$, $\epsilon$, $\lambda$, and $\mu$ are not assumed constant. Direct differentiation and a Legendre transform give
\begin{align}
&~~~~~~~~ \pi_\xi=2a^2\epsilon\left(\xi'-\frac{\mathcal H\lambda}{\sqrt{2\epsilon}}\psi\right),
 ~~~~~~~~~~~~~~~~\pi_\psi=a^2\psi',\label{eq:momenta}\\
 \mathcal H^{(2)}&=\frac{\pi_\xi^2}{4a^2\epsilon}+\frac{\pi_\psi^2}{2a^2}
 +a^2\epsilon(\nabla\xi)^2+\frac{a^2}{2}(\nabla\psi)^2
 +\frac{a^4\mu^2}{2}\psi^2
 +\frac{\mathcal H\lambda}{2\sqrt{2\epsilon}}
 (\pi_\xi\psi+\psi\pi_\xi).\label{eq:rawH}
\end{align}
Define $v_1=a\sqrt{2\epsilon}\,\xi$, $v_2=a\psi$, $Q_i=\sqrt{k}\,v_i$, and their canonically conjugate momenta $P_1=\pi_\xi/(a\sqrt{2\epsilon k})$, $P_2=\pi_\psi/(a\sqrt{k})$. The time-dependent transformation contributes the terms proportional to $\mathcal H$ and $\epsilon'/\epsilon$. For each real Fourier sector the complete Hamiltonian is
\begin{align}
 H_k&=\frac{k}{2}(P_1^2+P_2^2)+\frac{k}{2}Q_1^2
 +\frac12\left(k+\frac{a^2\mu^2}{k}\right)Q_2^2
 +\frac12\left(\mathcal H+\frac{\epsilon'}{2\epsilon}\right)(P_1Q_1+Q_1P_1)
 \nonumber\\
 &+\frac{\mathcal H\lambda}{2}(P_1Q_2+Q_2P_1)
 +\frac{\mathcal H}{2}(P_2Q_2+Q_2P_2).
 \label{eq:realH}
\end{align}
The derivation, including all noncommuting products, appears in Appendix~\ref{app:hamiltonian}. With $b_{i,\pm}=(Q_{i,\pm}+iP_{i,\pm})/\sqrt2$, the normal-ordered momentum-pair Hamiltonian is
\begin{align}
 \hat H_{\rm pair,k}& =k\left(n_{1,+}+n_{1,-}\right)
 +\left(k+\frac{a^2\mu^2}{2k}\right)\left(n_{2,+}+n_{2,-}\right)-\frac{i\mathcal H\lambda}{2}
 \left(b_{1,+}b_{2,-}+b_{2,+}b_{1,-}\right)
 \nonumber\\&+\frac{i\mathcal H\lambda}{2}\left(
 b_{1,+}^\dagger b_{2,+}+b_{1,-}^\dagger b_{2,-}
 -b_{2,+}^\dagger b_{1,+}-b_{2,-}^\dagger b_{1,-}\right)+\frac{i\mathcal H\lambda}{2}
 \left(b_{1,+}^\dagger b_{2,-}^\dagger+b_{2,+}^\dagger b_{1,-}^\dagger\right)
 \nonumber\\
 &-i\left(\mathcal H+\frac{\epsilon'}{2\epsilon}\right)b_{1,+}b_{1,-}
 +i\left(\mathcal H+\frac{\epsilon'}{2\epsilon}\right)b_{1,+}^\dagger b_{1,-}^\dagger+\left(\frac{a^2\mu^2}{2k}-i\mathcal H\right)b_{2,+}b_{2,-}
 \nonumber\\
 &
 +\left(\frac{a^2\mu^2}{2k}+i\mathcal H\right)b_{2,+}^\dagger b_{2,-}^\dagger
 +2k+\frac{a^2\mu^2}{2k}.
 \label{eq:pairH}
\end{align}
Here $n_{i,\pm}=b_{i,\pm}^\dagger b_{i,\pm}$. The displayed expression contains every same-field and cross-field pair terms, together with interfield transfer terms. Appendix~\ref{app:hamiltonian} also presents all eight coefficients obtained directly from the original coupled mode functions. To implement the four-mode squeezed state of Ref.~\cite{Colas:2021llj} without restricting its Fock space
\begin{align}
 |\Psi_k(\tau)\rangle={}&\mathcal N_k(\tau)
 \exp\!\left[Z_{11,k} b_{1,+}^\dagger b_{1,-}^\dagger
 +Z_{12,k} b_{1,+}^\dagger b_{2,-}^\dagger
 +Z_{12,k} b_{2,+}^\dagger b_{1,-}^\dagger
 +Z_{22,k} b_{2,+}^\dagger b_{2,-}^\dagger\right]|0_k\rangle,
 \label{eq:fourstate}\\
 |\mathcal N_k|^2={}&(1-|Z_{11,k}|^2-|Z_{12,k}|^2)
 (1-|Z_{22,k}|^2-|Z_{12,k}|^2)-|Z_{11,k}Z_{12,k}^*+Z_{12,k}Z_{22,k}^*|^2>0.
 \label{eq:normalization}
\end{align}
Normalizability also requires $1-|Z_{11,k}|^2-|Z_{12,k}|^2>0$. The three complex entries of the symmetric pair-amplitude matrix
\begin{equation}
 Z_k=\begin{pmatrix}Z_{11,k}&Z_{12,k}\\Z_{12,k}&Z_{22,k}\end{pmatrix},
 \label{eq:Zdefinition}
\end{equation}
where, $Z_{11,k}$ and $Z_{22,k}$ create same-field opposite-momentum pairs in the original field basis. The terms proportional to \(Z_{12,k}\) create the two cross-field pairs. Eq.~\eqref{eq:fourstate} is a compact Gaussian representation of the four-mode squeezed state discussed in Ref. \cite{Colas:2021llj}. Its relation to the occupation-number amplitudes is given in Appendix \ref{app:state}.

To express the complete state in terms of the two principal squeezing strengths and squeezing angles, we choose the passive transformation
\begin{equation}
 U_k=\begin{pmatrix}
 \cos\vartheta_k&-e^{i\chi_k}\sin\vartheta_k\\
 e^{-i\chi_k}\sin\vartheta_k&\cos\vartheta_k
 \end{pmatrix},\qquad
 Z_k=-U_k\begin{pmatrix}
 e^{2i\phi_{1k}}\tanh r_{1k}&0\\
 0&e^{2i\phi_{2k}}\tanh r_{2k}
 \end{pmatrix}U_k^T.
 \label{eq:Zsix}
\end{equation}
The six real parameters are $(r_{1k},r_{2k},\phi_{1k},\phi_{2k},\vartheta_k,\chi_k)$. The two squeezing strengths are obtained from the Takagi decomposition of the full pair-amplitude matrix. They characterize the principal squeezing channels of the joint four-mode squeezed state at each time. These channels need not coincide with the original curvature and entropy fields, nor do they imply a dynamical decoupling of the Hamiltonian. In particular, the entries in the original field basis are
\begin{align}
 Z_{11,k}={}&-\cos^2\vartheta_k\,e^{2i\phi_{1k}}\tanh r_{1k}
 -e^{2i\chi_k}\sin^2\vartheta_k\,e^{2i\phi_{2k}}\tanh r_{2k},\label{eq:Z11physical}\\
 Z_{12,k}={}&-\sin\vartheta_k\cos\vartheta_k\left[
 e^{-i\chi_k}e^{2i\phi_{1k}}\tanh r_{1k}
 -e^{i\chi_k}e^{2i\phi_{2k}}\tanh r_{2k}\right],\label{eq:Z12physical}\\
 Z_{22,k}={}&-e^{-2i\chi_k}\sin^2\vartheta_k\,e^{2i\phi_{1k}}\tanh r_{1k}
 -\cos^2\vartheta_k\,e^{2i\phi_{2k}}\tanh r_{2k}.
 \label{eq:Z22physical}
\end{align}
For a pair-amplitude matrix diagonal in the chosen field basis, \(\vartheta_k=0\) may be chosen and \(Z_{12,k}=0\). The dynamically uncoupled model is obtained by setting the interfield coupling to zero and evolving its equations separately. Each diagonal entry then becomes $-e^{2i\phi_{ik}}\tanh r_{ik}$, and the cross amplitude vanishes. With coupling, the phases of the three matrix entries generally differ from $\phi_{1k}$ and $\phi_{2k}$.

Acting with the four annihilation operators on Eq.~\eqref{eq:fourstate} gives, for example
\begin{align}
 b_{1,+}|\Psi_k\rangle&=(Z_{11,k}b_{1,-}^\dagger+Z_{12,k}b_{2,-}^\dagger)|\Psi_k\rangle,
 &b_{2,+}|\Psi_k\rangle&=(Z_{12,k}b_{1,-}^\dagger+Z_{22,k}b_{2,-}^\dagger)|\Psi_k\rangle,\nonumber\\
 b_{1,-}|\Psi_k\rangle&=(Z_{11,k}b_{1,+}^\dagger+Z_{12,k}b_{2,+}^\dagger)|\Psi_k\rangle,
 &b_{2,-}|\Psi_k\rangle&=(Z_{12,k}b_{1,+}^\dagger+Z_{22,k}b_{2,+}^\dagger)|\Psi_k\rangle.
 \label{eq:annihilation}
\end{align}
Substitution of these identities and Eq.~\eqref{eq:pairH} into $i\partial_\tau|\Psi_k\rangle=\hat H_{\rm pair,k}|\Psi_k\rangle$ produces three coupled complex differential equations
\begin{equation}
    \begin{split}
         iZ_{11,k}'&=i\left(\mathcal H+\frac{\epsilon'}{2\epsilon}\right)
 +2kZ_{11,k}+i\mathcal H\lambda Z_{12,k}
  -i\left(\mathcal H+\frac{\epsilon'}{2\epsilon}\right)Z_{11,k}^2 -i\mathcal H\lambda Z_{11,k}Z_{12,k}\\&
 +\left(\frac{a^2\mu^2}{2k}-i\mathcal H\right)Z_{12,k}^2,
 \label{eq:xODE}
    \end{split}
\end{equation}
\begin{equation}
    \begin{split}
         iZ_{12,k}'&=\frac{i\mathcal H\lambda}{2}
 +\left(2k+\frac{a^2\mu^2}{2k}\right)Z_{12,k}
 +\frac{i\mathcal H\lambda}{2}(Z_{22,k}-Z_{11,k}) -i\left(\mathcal H+\frac{\epsilon'}{2\epsilon}\right)Z_{11,k}Z_{12,k}
\\&
 -\frac{i\mathcal H\lambda}{2}(Z_{11,k}Z_{22,k}+Z_{12,k}^2)
 +\left(\frac{a^2\mu^2}{2k}-i\mathcal H\right)Z_{12,k}Z_{22,k},
 \label{eq:yODE}
    \end{split}
\end{equation}
\begin{equation}
    \begin{split}
         iZ_{22,k}'&=\frac{a^2\mu^2}{2k}
 +2\left(k+\frac{a^2\mu^2}{2k}\right)Z_{22,k}
-i\left(\mathcal H+\frac{\epsilon'}{2\epsilon}\right)Z_{12,k}^2-i\mathcal H\lambda \big(Z_{12,k}Z_{22,k}+ Z_{12,k}\big)\\&
 +\left(\frac{a^2\mu^2}{2k}-i\mathcal H\right)Z_{22,k}^2.
 \label{eq:zODE}
    \end{split}
\end{equation}
For the initial four-mode vacuum state, $Z_{11,k}(\tau_i)=Z_{12,k}(\tau_i)=Z_{22,k}(\tau_i)=0$, $Z_{12,k}'(\tau_i)=\mathcal H(\tau_i)\lambda(\tau_i)/2$. Thus, a nonzero turn rate sources the initial growth of the cross-field pair amplitude, although the initial state itself is uncorrelated. For the numerical illustration, a leading-order quasi-de Sitter background is used with $a(\tau)=-1/(H_0\tau)$ and $\epsilon(\tau)=\epsilon_{\rm ref}(-k\tau)^{-\eta_{\rm SR}}$. The parameters are $H_0=1$, $\epsilon_{\rm ref}=0.002$, $\eta_{\rm SR}=0.01$, $\mu=0.5H_0$, $\lambda=0.1$, and $k=0.01$. Evolution starts at $\tau_i=-10000$ and ends at $\tau_f=-0.02$. The uncoupled comparison sets $\lambda=0$ while retaining the same prescribed background, effective mass, initial vacuum, and circuit cost.

\begin{figure}[!h]
    \centering
    \includegraphics[width=0.55\linewidth]{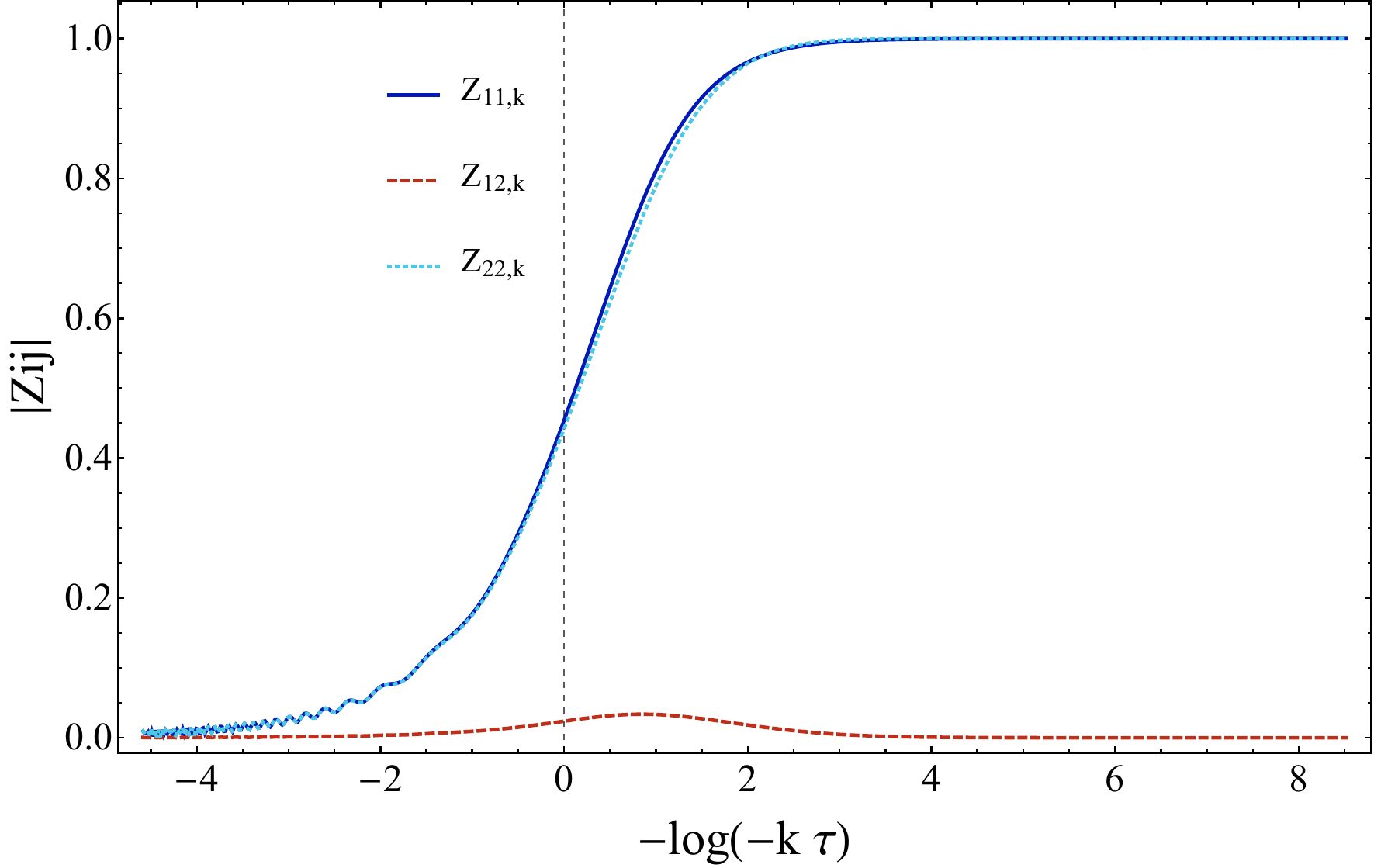}
    \includegraphics[width=0.47\linewidth]{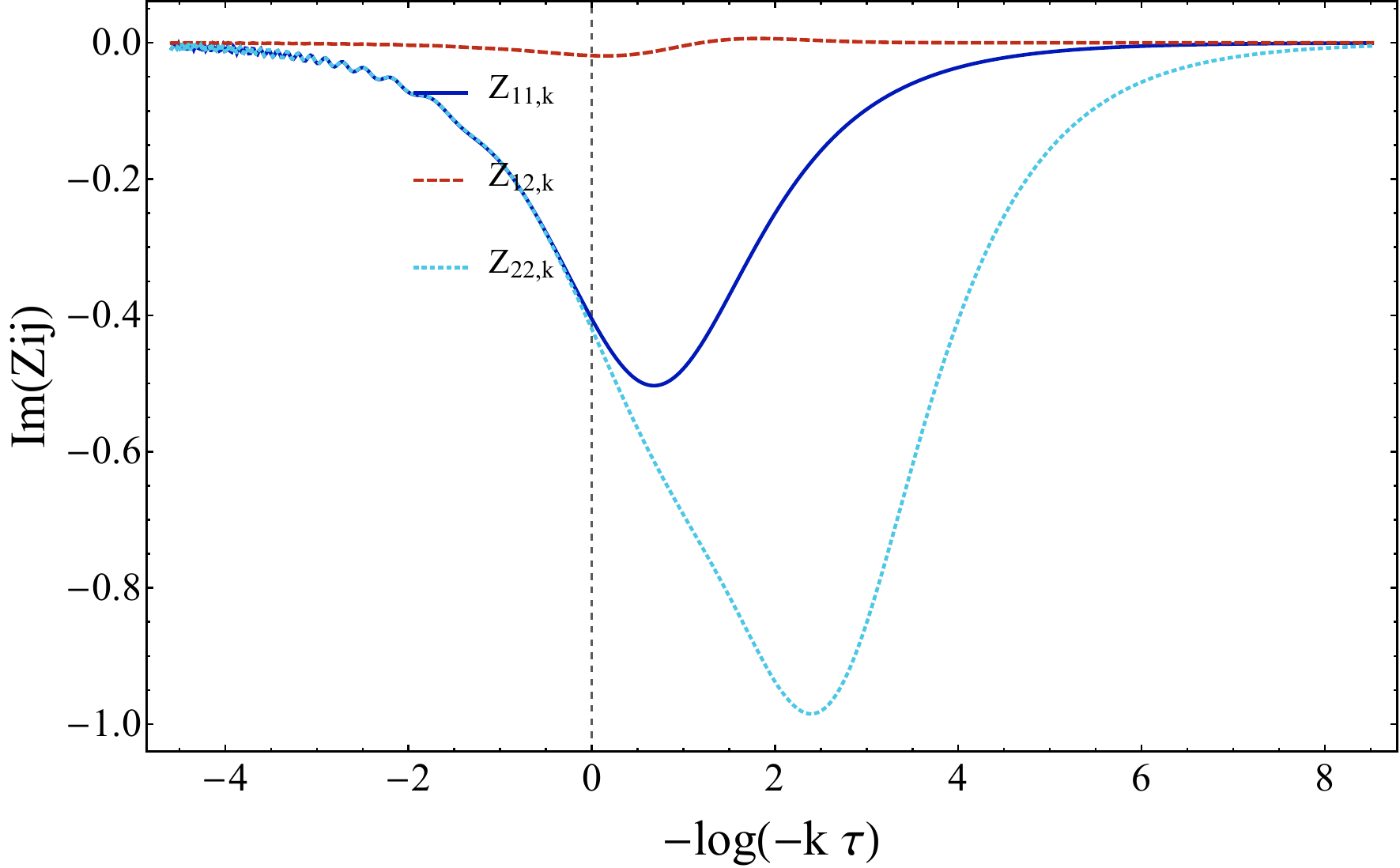}
    \includegraphics[width=0.47\linewidth]{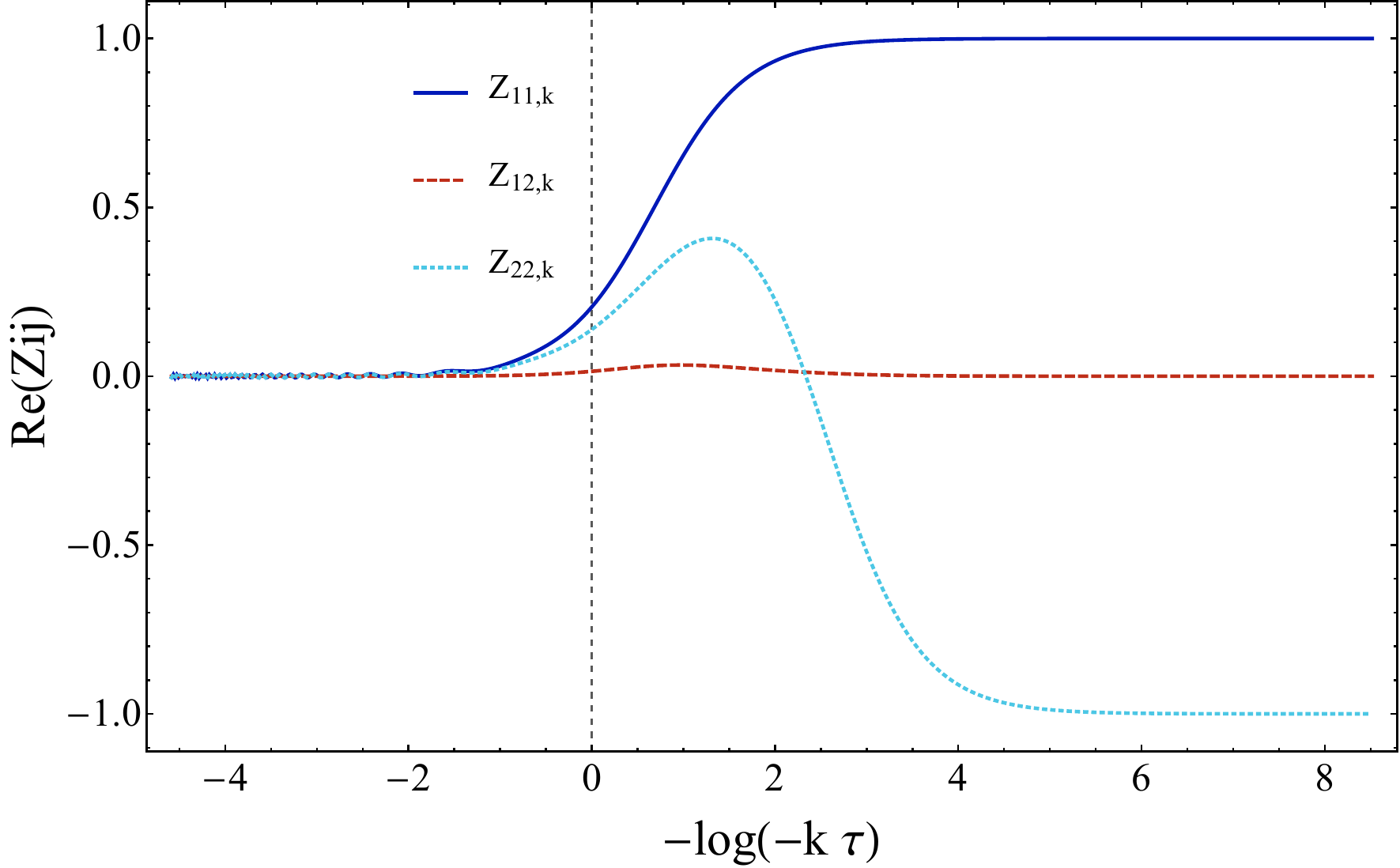}
    \caption{Evolution of the four-mode pair amplitudes \(Z_{11,k}\), \(Z_{12,k}\), and \(Z_{22,k}\) in the coupled two-field inflationary model. The upper panel shows their moduli. The lower-left and lower-right panels show their imaginary and real parts, respectively. The horizontal coordinate is \(x=-\log(-k\tau)\), and the vertical dashed line marks horizon crossing at \(x=0\). The diagonal entries describe opposite-momentum pairs within the original field basis, while \(Z_{12,k}=Z_{21,k}\) describes cross-field pairs.}
    \label{fig:amplitudes}
\end{figure}

Fig. \ref{fig:amplitudes} shows how the four-mode pair amplitudes evolve in the original field basis. Deep inside the horizon, the mode frequency dominates the background-dependent terms in the quadratic Hamiltonian, and all three amplitudes remain small. As the mode approaches horizon crossing, the relative importance of the background terms increases. The diagonal amplitudes grow, while the interfield amplitude develops a smaller transient enhancement. Its subsequent decrease does not mean that the earlier coupling has ceased to affect the state: the diagonal amplitudes have already evolved under the coupled equations. Moreover, the principal squeezing strengths depend on the eigenvalues of the full pair-amplitude matrix rather than on the instantaneous magnitude of the interfield entry alone. The different real and imaginary parts of the diagonal amplitudes also show that phase information must be retained when those strengths and their angles are extracted.

The overall normalization and phase satisfy
\begin{equation}
 i\frac{\mathcal N_k'}{\mathcal N_k}=2k+\frac{a^2\mu^2}{2k}
 -i\left(\mathcal H+\frac{\epsilon'}{2\epsilon}\right)Z_{11,k}
 -i\mathcal H\lambda Z_{12,k}
 +\left(\frac{a^2\mu^2}{2k}-i\mathcal H\right)Z_{22,k}.
 \label{eq:NODE}
\end{equation}
The two independent nonnegative squeezing strengths of the complete state follow from the explicit quadratic equation derived in Appendix~\ref{app:state}:
{\small\begin{align}
 r_{1k}={}&\operatorname{arctanh}\!\sqrt{\frac{|Z_{11,k}|^2+|Z_{22,k}|^2+2|Z_{12,k}|^2
 +\sqrt{(|Z_{11,k}|^2-|Z_{22,k}|^2)^2+4|Z_{11,k}Z_{12,k}^*+Z_{12,k}Z_{22,k}^*|^2}}{2}},
 \label{eq:r1}\\
 r_{2k}={}&\operatorname{arctanh}\!\sqrt{\frac{|Z_{11,k}|^2+|Z_{22,k}|^2+2|Z_{12,k}|^2
 -\sqrt{(|Z_{11,k}|^2-|Z_{22,k}|^2)^2+4|Z_{11,k}Z_{12,k}^*+Z_{12,k}Z_{22,k}^*|^2}}{2}}.
 \label{eq:r2}
\end{align}}
Eqs.~\eqref{eq:r1} and \eqref{eq:r2} convert the three complex pair amplitudes into the two principal squeezing strengths. The squeezing angles are obtained in the same phase convention
from \(\phi_{ik}=\tfrac12\arg(-w_{ik})\pmod{\pi}\), where the rotated pair amplitudes \(w_{1k}\) and \(w_{2k}\) are given
explicitly in Eqs.~\eqref{eq:w1}--\eqref{eq:phaseextract}.

\begin{figure}[!h]
    \centering
    \includegraphics[width=0.47\linewidth]{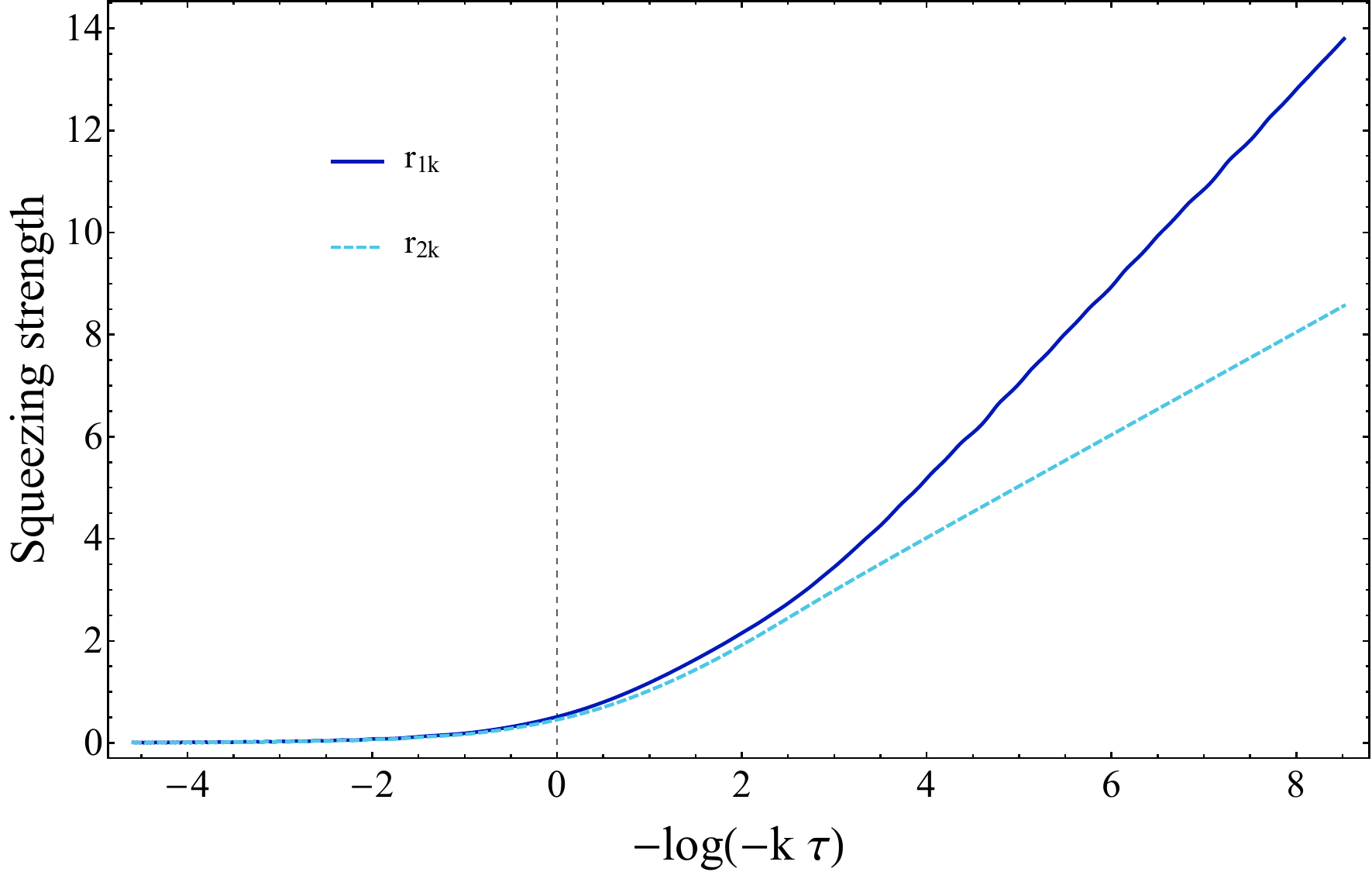}
    \includegraphics[width=0.47\linewidth]{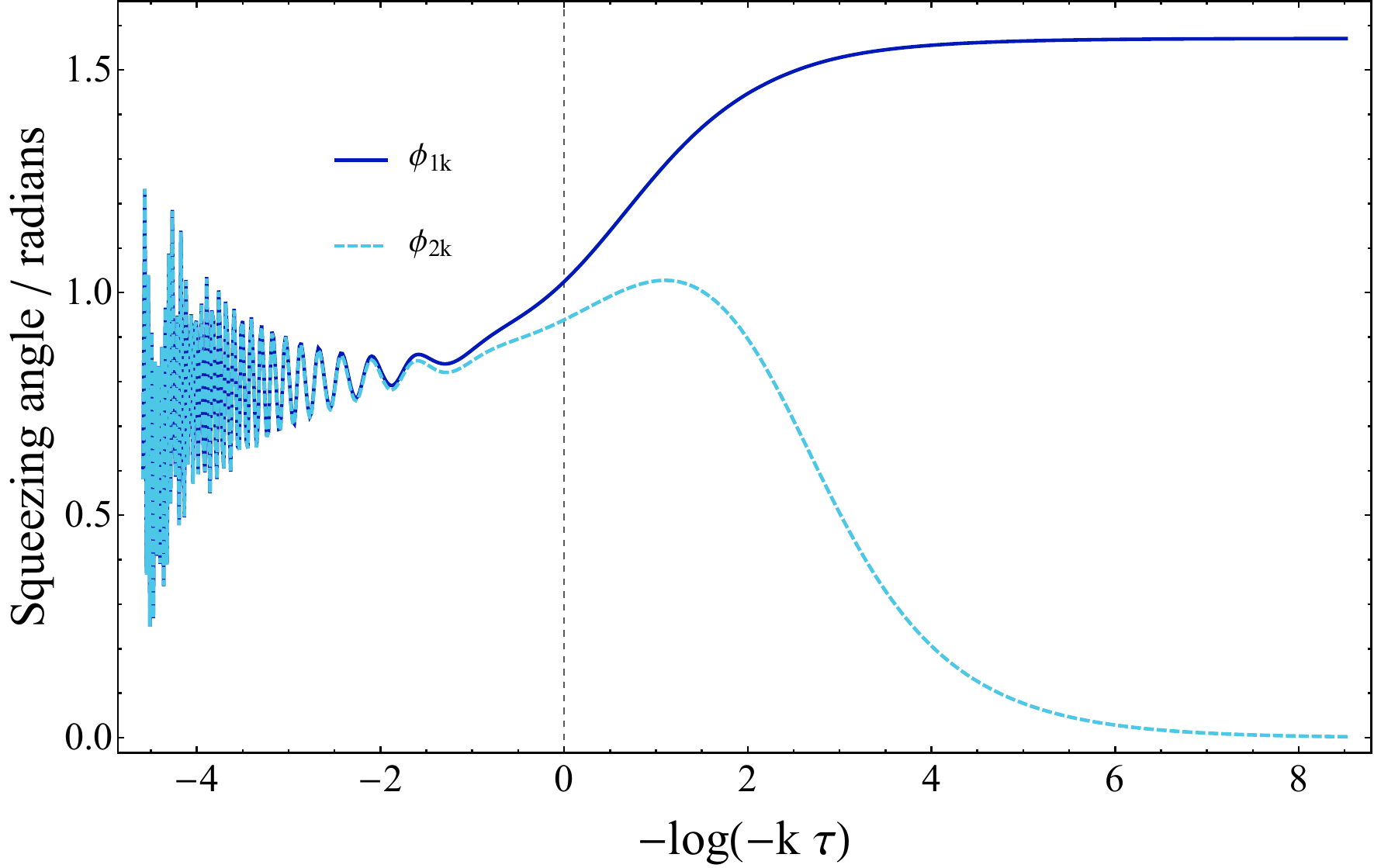}
    \caption{Evolution of the principal squeezing strengths \(r_{1k}\) and \(r_{2k}\) (left) and squeezing angles \(\phi_{1k}\) and \(\phi_{2k}\) (right) in the coupled two-field inflationary model. The horizontal coordinate is \(x=-\log(-k\tau)\), and the vertical dashed line marks horizon crossing, \(x=0\). The strengths and angles are extracted from the complete four-mode pair-amplitude matrix, including its cross-field entry \(Z_{12,k}\). The ordered strengths \(r_{1k}\geq r_{2k}\) label principal squeezing channels rather than the original curvature and entropy fields.}
    \label{fig:squeezed parameter}
\end{figure}

Fig.~\ref{fig:squeezed parameter} shows the principal squeezing strengths extracted from the complete pair-amplitude matrix. Both squeezing strengths are small in the subhorizon regime and increase after horizon crossing. This behavior reflects the amplification of inflationary perturbations as the background-dependent terms become important relative to the mode frequency. The two squeezing strengths separate at late times because the coupled curvature and entropy sectors need not be amplified equally. These are principal squeezing channels of the joint state, not permanent labels for the original fields. The squeezing angles retain information about the phases of the evolving Gaussian state. Their early variations should be interpreted with care because an angle associated with a nearly vanishing squeezing strength is poorly determined. The later angular behavior depends on the phase convention. For the equal-weight circuit geometry used below, the angles have no independent contribution to the minimum distance when the two principal squeezing strengths are held fixed. They can nevertheless influence that distance indirectly through the phase-dependent evolution of the full state.

The two squeezing strengths are ordered as $r_{1k}\geq r_{2k}\geq0$, without identification with either original field. Combining their explicitly evaluated values with the circuit distance yields
\begin{equation}
 \mathcal C_{2,k}(r_{1k},r_{2k},\phi_{1k},\phi_{2k},\vartheta_k,\chi_k)
 =\sqrt{2\left(r_{1k}^2+r_{2k}^2\right)}.
 \label{eq:Cexplicit}
\end{equation}
There is no single universal $r_k$ for a coupled four-mode squeezed state. The $r_{1k}$ and $r_{2k}$ are its two principal squeezing strengths. When only one channel is squeezed, $r_{2k}=0$ and $\mathcal C_{2,k}=\sqrt2\,r_{1k}$. When both have the same strength $r_{1k}=r_{2k}=r_k$, $\mathcal C_{2,k}=2r_k$. In the decoupled limit the two strengths are the separate one-field $r_k$ values, up to ordering. Eq. \eqref{eq:anglederivatives} does not enter the equal-weight circuit complexity explicitly. The squeezing and mixing angles are extracted from the same coupled amplitudes \(Z_{11,k},Z_{12,k},Z_{22,k}\) that determine the squeezing strengths. Relative phases of these amplitudes contribute to the interference term \(|Z_{11,k}Z_{12,k}^{*}+Z_{12,k}Z_{22,k}^{*}|^2\), and can thereby modify \(r_{1k}\) and \(r_{2k}\). Their effect on the present complexity is therefore indirect. At fixed principal squeezing strengths, changes of the squeezing angles alone leave \(\mathcal C_{2,k}\) invariant. No expression for arbitrary time-dependent $a,\epsilon,\lambda,\mu$ can eliminate the initial-value problem in Eqs.~\eqref{eq:xODE}--\eqref{eq:zODE}. Eqs.~\eqref{eq:r1}--\eqref{eq:Cexplicit} are explicit algebraic functions of its three complex solutions. In the uncoupled limit $Z_{12,k}=0$, the two strengths reduce to the ordered values of $\operatorname{arctanh}|Z_{11,k}|$ and $\operatorname{arctanh}|Z_{22,k}|$. The equal-weight circuit complexity is evaluated from the two principal squeezing strengths in each evolution. 

\begin{figure}[!h]
    \centering
    \includegraphics[width=0.48\linewidth]{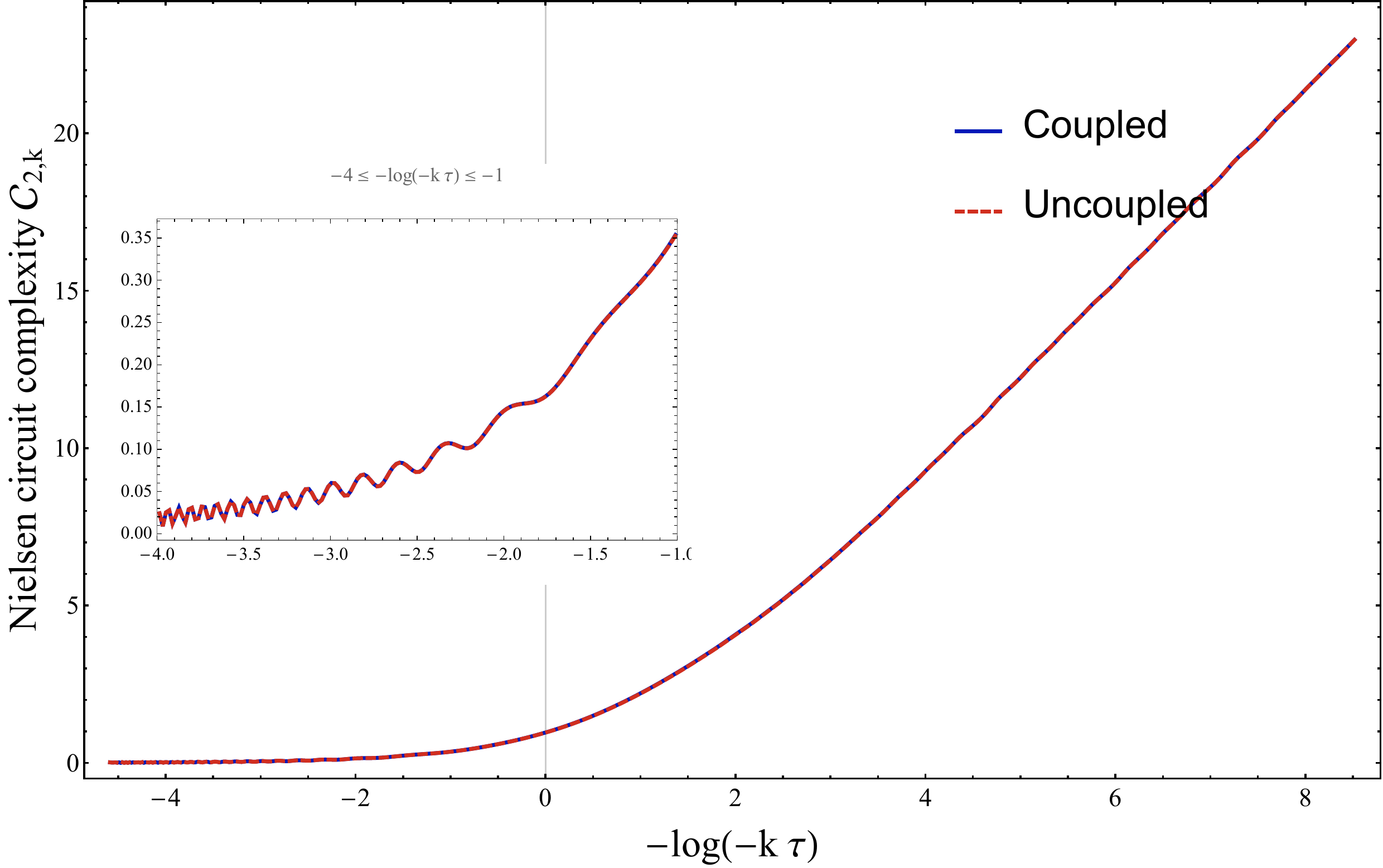}
       \includegraphics[width=0.48\linewidth]{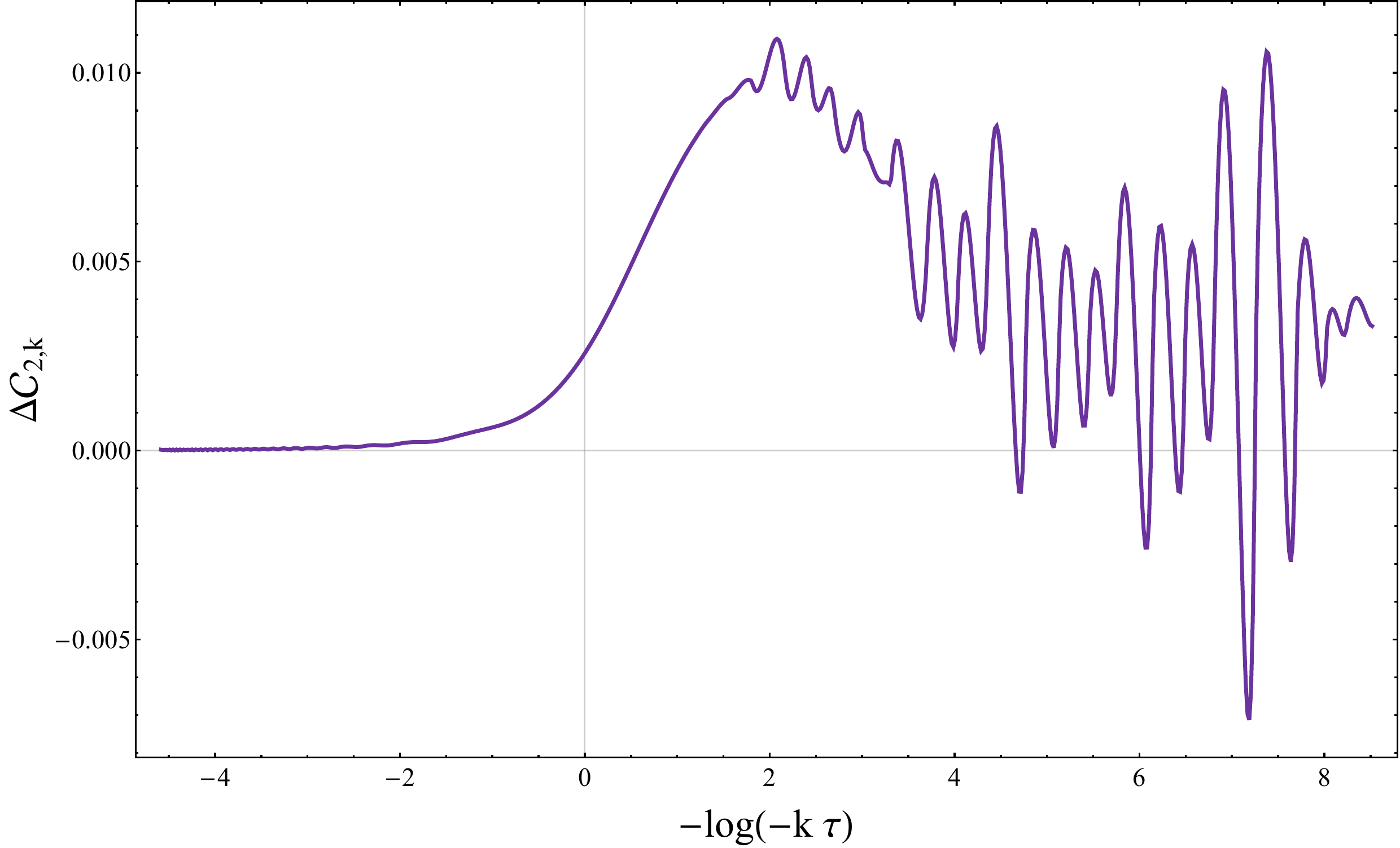}
    \caption{Equal-weight Nielsen $F_2$ circuit complexity of one complete
momentum pair in two-field inflation. The left panel compares the
coupled evolution with the matched uncoupled evolution obtained by
setting $\lambda=0$ while keeping the background, mass, initial vacuum,
and circuit convention fixed. The inset enlarges
$-4\leq-\log(-k\tau)\leq-1$. The right panel shows the signed difference
$\Delta\mathcal C_{2,k}
=\mathcal C_{2,k}^{\rm coupled}
-\mathcal C_{2,k}^{\lambda=0}$.
The vertical dashed line marks approximate horizon crossing at \(-\log(-k\tau)=0\). }
    \label{fig:cc}
\end{figure}

Fig.~\ref{fig:cc} compares the coupled and matched uncoupled circuit complexities and their signed difference. Both curves show early oscillations and grow after horizon crossing. Their common growth reflects inflationary squeezing in both evolutions. The signed difference is much smaller than either full complexity and is resolved on the narrower scale of the right panel. Neither the similar growth of the curves nor the onset of a visible difference determines a Lyapunov exponent or a scrambling time. These
quantities require a separately defined sensitivity diagnostic. The coupled evolution changes both the cross-field and same-field pair amplitudes, so its pair-amplitude matrix differs from that of the matched uncoupled evolution. Nevertheless, the equal-weight distance depends only on the two principal squeezing strengths
and has no separate dependence on the orientation of their channels at fixed strengths \cite{Chapman:2018hou}. For the parameters considered, the resulting preparation distances are nearly equal. Thus, this particular complexity does not resolve every change in the four-mode squeezed state. The quadratic model also satisfies\(\mathcal C_{2,k}(\lambda)=\mathcal C_{2,k}(-\lambda)\). Reversing the signs of the entropy field and its conjugate momentum maps one coupling sign to the other. This transformation preserves the vacuum reference and the equal-weight cost. Where the complexity is smooth at \(\lambda=0\), its linear correction therefore vanishes. The magnitude of the difference at the finite coupling used here is determined by the coupled and uncoupled numerical evolutions.

\section{Conclusion and outlook}
\label{sec:conclusion}

In this work, we evaluate the Nielsen circuit complexity of coupled curvature and entropy perturbations using the four-mode squeezed state. The quadratic Hamiltonian includes both diagonal and off-diagonal entries of $A_{ij,k}$ and $B_{ij,k}$. Their combined evolution determines the three complex pair amplitudes $Z_{11,k}$, $Z_{12,k}$ and $Z_{22,k}$. The eigenvalues of \(Z_kZ_k^\dagger\) determine the two principal squeezing strengths of the joint state. These strengths cannot generally be assigned separately to the original curvature and entropy fields.

For the vacuum reference and the equal-weight Gaussian circuit, the minimum distance is determined by the two principal squeezing strengths of the complete state. The matched coupled and uncoupled evolutions exhibit nearly identical complexity curves for the parameters considered. Both grow after horizon crossing because inflationary amplification squeezes the opposite-momentum modes in either model. The coupling generates an interfield pair amplitude and changes the joint Gaussian state, but its effect on this particular preparation distance is small. The signed difference makes that effect visible only on a much narrower scale. This result illustrates the dependence of a complexity comparison on its reference state, gate set, and cost function. It does not imply that interfield correlations are absent or dynamically irrelevant.

Unequal gate penalties can, in principle, make the preparation cost sensitive to the orientation and angular evolution of the four-mode squeezed state. The weighted results presented here give bounds and a trial path. An exact weighted geodesic for coupled channels remains to be found. Although the complexity grows after horizon crossing, the present calculation does not determine a Lyapunov exponent or a scrambling time. Establishing either quantity requires a separately defined growth or sensitivity diagnostic. It would also be useful to compare this Nielsen diagnostic directly with the Krylov observables studied in two-field inflation, while keeping their distinct definitions explicit \cite{Li:2026cry}. Circuit complexity in the present work characterizes state preparation within a chosen Gaussian geometry. Its connection to primordial observables requires a separate analysis.

\appendix
\section{Four-mode wavefunction and the minimum circuit length}
\label{app:covariance}

Set $b_{i,c}=(b_{i,+}+b_{i,-})/\sqrt2$ and $b_{i,s}=(b_{i,+}-b_{i,-})/(i\sqrt2)$. Commutativity of the creation operators yields
\begin{align}
 &~~~~~~~~~~~~~~Z_{11,k} b_{1,+}^\dagger b_{1,-}^\dagger
 +Z_{12,k}(b_{1,+}^\dagger b_{2,-}^\dagger+b_{2,+}^\dagger b_{1,-}^\dagger)
 +Z_{22,k} b_{2,+}^\dagger b_{2,-}^\dagger\nonumber\\
 &\quad=\frac{Z_{11,k}}{2}\bigl((b_{1,c}^\dagger)^2+(b_{1,s}^\dagger)^2\bigr)
 +Z_{12,k}\bigl(b_{1,c}^\dagger b_{2,c}^\dagger+b_{1,s}^\dagger b_{2,s}^\dagger\bigr)
 +\frac{Z_{22,k}}{2}\bigl((b_{2,c}^\dagger)^2+(b_{2,s}^\dagger)^2\bigr).
 \label{eq:cs}
\end{align}
To see the cancellation explicitly, invert the definitions of the real-sector operators
\begin{equation}
    b_{i,+}=\frac{b_{i,c}+i b_{i,s}}{\sqrt{2}},
\qquad
b_{i,-}=\frac{b_{i,c}-i b_{i,s}}{\sqrt{2}}.
\end{equation}
Substituting these relations into the exponent of Eq.~\eqref{eq:fourstate}, the terms containing one cosine and one sine creation operator cancel. This cancellation uses \(Z_{12,k}=Z_{21,k}\) and the commutativity of the creation operators. The remaining exponent is
\begin{equation}
    \frac12 b_c^\dagger Z_kb_c^{\dagger T}+\frac12 b_s^\dagger Z_kb_s^{\dagger T}.
\end{equation}
Thus the four-mode wavefunction consists of two identical real Gaussian sectors. Both sectors must be included in the trace that defines the complexity of the complete momentum pair. For either real sector, the state satisfies
\begin{equation}
    \left( b_\nu-Z_k b_\nu^\dagger\right)|\Psi_k\rangle=0,\qquad \nu=c,s.
\end{equation}
In the coordinate representation, \( b_\nu=( Q_\nu+\partial_\nu)/\sqrt{2}\) and \( b_\nu^\dagger=( Q_\nu-\partial_\nu)/\sqrt{2}\). The annihilation equation therefore becomes
\begin{equation}
    \left[(I_2-Z_k)\boldsymbol Q_\nu +(I_2+Z_k)\boldsymbol\partial_\nu\right]\Psi_k=0.
\end{equation}
Writing \(\Psi_k\propto
\exp[-\boldsymbol Q_\nu^{T}W\boldsymbol Q_\nu/2]\) in the annihilation equation gives \(W=(I_2+Z_k)^{-1}(I_2-Z_k)\). The two matrix factors commute because both are functions of \(Z_k\). Evaluating the inverse gives the following entries of the Gaussian kernel
\begin{equation}
 \begin{aligned}
 W_{11}&=\frac{(1-Z_{11,k})(1+Z_{22,k})+Z_{12,k}^2}
 {(1+Z_{11,k})(1+Z_{22,k})-Z_{12,k}^2},\\
 W_{12}=W_{21}&=\frac{-2Z_{12,k}}
 {(1+Z_{11,k})(1+Z_{22,k})-Z_{12,k}^2},\\
 W_{22}&=\frac{(1-Z_{22,k})(1+Z_{11,k})+Z_{12,k}^2}
 {(1+Z_{11,k})(1+Z_{22,k})-Z_{12,k}^2}.
 \end{aligned}
 \label{eq:Wexplicit}
\end{equation}
Writing $W=(W_{ij})=X+iY$, where $X$ and $Y$ are real symmetric matrices and $X$ is positive definite, the full normalized four-mode wavefunction reads
\begin{equation}
 \Psi( Q_c,Q_s)=\frac{\sqrt{\det X}}{\pi} \exp\!\left[-\frac12 Q_c^T(X+iY) Q_c
-\frac12 Q_s^T(X+iY) Q_s\right].
 \label{eq:wavefunction}
\end{equation}
For either real sector, the probability density derived from Eq.~\eqref{eq:wavefunction} is proportional to \(\exp(- Q_\nu^T X Q_\nu)\). Gaussian integration therefore gives \(\langle Q_\nu Q_\nu^T\rangle =X^{-1}/2\). Moreover, \(P_i=-i\partial/\partial Q_i\) and Eq.~\eqref{eq:wavefunction} imply \( P_\nu\Psi=i(X+iY) Q_\nu\Psi\). The symmetrized position--momentum correlator is \(-X^{-1}Y/2\), while the momentum correlator is \((X+YX^{-1}Y)/2\). Combining these three blocks gives
\begin{equation}
 G_{T,k}=\frac12\begin{pmatrix}
 X^{-1}&-X^{-1}Y\\-YX^{-1}&X+YX^{-1}Y
 \end{pmatrix}.
 \label{eq:covariance}
\end{equation}
We now connect this target covariance to the circuit definition in Eq.~\eqref{eq:cost}. The vacuum reference has covariance \(G_R=I_8/2\). If a Gaussian circuit acts on the quadratures by the symplectic matrix \(S(s)\), its endpoint prepares the target state precisely when
\begin{equation}
    S(0)=I_8,\qquad S(1)S(1)^T=2G_T,\qquad G_T=G_{T,k}\oplus G_{T,k}.
\end{equation}
The endpoint condition fixes the target covariance, but does not select a unique endpoint matrix: a right orthogonal symplectic factor leaves \(S(1)S(1)^T\) unchanged. The two strengths \(r_{1k}\) and \(r_{2k}\) are obtained from the singular values of the complete pair-amplitude matrix:
\begin{equation}
    \operatorname{spec}(Z_kZ_k^\dagger) = \{\tanh^2r_{1k},\tanh^2r_{2k}\}.
\end{equation}
A passive transformation identifies these principal squeezing channels without changing the vacuum. In one real sector, their quadrature squeezing factors are \(e^{r_{1k}},e^{r_{2k}},e^{-r_{1k}},e^{-r_{2k}}\). The corresponding orthogonal symplectic transformations change the principal axes of the covariance, but not its eigenvalues. It follows that
\begin{equation}
    \operatorname{spec}(2G_{T,k}) = \{e^{2r_{1k}},e^{2r_{2k}},  e^{-2r_{1k}},
    e^{-2r_{2k}}\}.
\end{equation}
Every eigenvalue occurs twice in \(2G_T\), because the cosine and sine sectors have identical covariances. A circuit ending at its positive square root has constant velocity $\frac12\log[2(G_{T,k}\oplus G_{T,k})]$. Its squared length is $\frac18\text{Tr}_8\{\log[2(G_{T,k}\oplus G_{T,k})]\}^2=2(r_{1k}^2+r_{2k}^2)$. For any other allowed circuit, write the logarithms of its singular values as $\ell_j(s)$. The singular-value variation bound gives
\begin{equation}
 \frac1{\sqrt2}\int_0^1ds\,\sqrt{\sum_{j=1}^8\dot\ell_j(s)^2}
 \leq\int_0^1ds\,\sqrt{\frac12\text{Tr}_8[(\dot{\mathbb S}\mathbb S^{-1})(\dot{\mathbb S}\mathbb S^{-1})^T]},
 \label{eq:bound}
\end{equation}
Since \(S(0)=I_8\), all initial logarithmic singular values vanish. Applying the triangle inequality to the vector
\(\boldsymbol\ell(s)=(\ell_1(s),\ldots,\ell_8(s))\) gives, for every admissible circuit,
\begin{equation}
    L_{F_2}[S]\geq\frac{1}{\sqrt{2}}\int_0^1 ds\,\|\dot{\boldsymbol\ell}(s)\|_2\geq\frac{1}{\sqrt{2}}\|\boldsymbol\ell(1)\|_2 .
\end{equation}
At the endpoint, the singular values are \(e^{r_{1k}},e^{-r_{1k}},e^{r_{2k}},e^{-r_{2k}}\), each occurring twice. Hence \(\|\boldsymbol\ell(1)\|_2^2
=4(r_{1k}^2+r_{2k}^2)\), and every circuit has length at least \(\sqrt{2(r_{1k}^2+r_{2k}^2)}\). The positive-square-root path attains this bound,
which proves Eq.~\eqref{eq:Crr}. 

For the weighted extension, normalize every generator by $\tfrac12\text{Tr}_8[(\mathbb J\mathsf H_I)(\mathbb J\mathsf H_J)^T]=\delta_{IJ}$ and expand $\tfrac12\log(2\mathbb G_T)$ in this basis. Because $w_I\geq w_{\min}$, the length of every admissible circuit exceeds $w_{\min}$ times its equal-weight length. The radial circuit gives the upper bound in Eq.~\eqref{eq:weightedbounds}. In the decoupled case its generator, in the $(Q_i,P_i)$ plane of each real sector
\begin{equation}
 K_{i,\rm rad}=-r_{ik}\begin{pmatrix}
 \cos(2\phi_{ik})&\sin(2\phi_{ik})\\
 \sin(2\phi_{ik})&-\cos(2\phi_{ik})
 \end{pmatrix}.
 \label{eq:radialgenerator}
\end{equation}
The same block occurs in the cosine and sine sectors. Squaring the diagonal and off-diagonal entries, applying their respective weights, and including both real sectors yields Eq.~\eqref{eq:weightedradial}. With $w_D=w_X=1$, it reduces to Eq.~\eqref{eq:Crr}. For $w_D\ne w_X$, the path length carries the angles $\phi_{ik}$, while its minimum over all paths must still be determined from Eq.~\eqref{eq:Nielsenweighted}.

Once a complete quadratic gate basis and its weights have been fixed, an exact numerical Nielsen extension can be approached by discretizing the circuit coordinate as $s_n=n/N$ and minimizing over the real controls $Y_n^I$
\begin{equation}
 \begin{aligned}
 \mathbb S_{n+1}&=\exp\!\left[\frac1N\sum_IY_n^I\mathbb J\mathsf H_I\right]\mathbb S_n,
 \qquad \mathbb S_0=I_8,\\
 \mathcal C_{\boldsymbol w,k}^{(N)}&=\min_{\{Y_n^I\}}\frac1N\sum_{n=0}^{N-1}
 \sqrt{\sum_I w_I^2(Y_n^I)^2},\qquad
 \mathbb S_N\mathbb S_N^T=2\mathbb G_T.
 \end{aligned}
 \label{eq:discreteNielsen}
\end{equation}
Increasing $N$ and checking convergence tests the numerical minimization. The endpoint is the covariance constructed from the numerically evolved $Z_{11,k},Z_{12,k},Z_{22,k}$. The physical inflationary trajectory is not imposed on the circuit path.

\section{Derivation of all Hamiltonian coefficients}
\label{app:hamiltonian}

\subsection{Canonical transformation and fixed vacuum}
Substitute $\xi=v_1/(a\sqrt{2\epsilon})$ and $\psi=v_2/a$ into Eq.~\eqref{eq:action}. The Lagrangian for one real Fourier sector becomes
\begin{align}
 L_k={}&\frac12\left[v_1'-\left(\mathcal H+\frac{\epsilon'}{2\epsilon}\right)v_1
 -\mathcal H\lambda v_2\right]^2
 +\frac12(v_2'-\mathcal H v_2)^2
 -\frac{k^2}{2}v_1^2-\frac{k^2+a^2\mu^2}{2}v_2^2.
 \label{eq:lagr}
\end{align}
Its momenta are $p_1=v_1'-(\mathcal H+\epsilon'/2\epsilon)v_1-\mathcal H\lambda v_2$ and $p_2=v_2'-\mathcal H v_2$. Inserting $v_i=Q_i/\sqrt{k}$, $p_i=\sqrt{k}P_i$ in $p_1v_1'+p_2v_2'-L_k$ gives Eq.~\eqref{eq:realH}. Using $Q_i=(b_i+b_i^\dagger)/\sqrt2$ and $P_i=(b_i-b_i^\dagger)/(i\sqrt2)$, the coefficients multiplying $b_{i,+}b_{j,-}$ are explicitly
\begin{equation}
 \begin{aligned}
 \mathcal A_{11}&=-i\left(\mathcal H+\frac{\epsilon'}{2\epsilon}\right),&
 \mathcal A_{12}&=-\frac{i\mathcal H\lambda}{2},\\
 \mathcal A_{21}&=-\frac{i\mathcal H\lambda}{2},&
 \mathcal A_{22}&=\frac{a^2\mu^2}{2k}-i\mathcal H,
 \end{aligned}
 \label{eq:Aentries}
\end{equation}
and those multiplying $b_{i,\pm}^\dagger b_{j,\pm}$ are
\begin{equation}
 \begin{aligned}
 \mathcal B_{11}&=k,&\mathcal B_{12}&=\frac{i\mathcal H\lambda}{2},\\
 \mathcal B_{21}&=-\frac{i\mathcal H\lambda}{2},&
 \mathcal B_{22}&=k+\frac{a^2\mu^2}{2k}.
 \end{aligned}
 \label{eq:Bentries}
\end{equation}
The pair-creation coefficients are the complex conjugates of Eq.~\eqref{eq:Aentries}. These explicit entries reproduce every term in Eq.~\eqref{eq:pairH}, including the normal-ordering constant $\mathcal B_{11}+\mathcal B_{22}$.

\subsection{All entries in the original coupled-mode-function basis}
If the original perturbations have been evolved as $\hat\xi_{\bm k}=\xi_1a_{1,\bm k}+\xi_2a_{2,\bm k}+\xi_1^*a_{1,-\bm k}^\dagger+\xi_2^*a_{2,-\bm k}^\dagger$ and $\hat\psi_{\bm k}=\psi_1a_{1,\bm k}+\psi_2a_{2,\bm k}+\psi_1^*a_{1,-\bm k}^\dagger+\psi_2^*a_{2,-\bm k}^\dagger$, their momenta follow from Eq.~\eqref{eq:momenta}. Direct substitution into Eq.~\eqref{eq:rawH} gives every pair-annihilation coefficient in terms of the physical mode functions and their canonical momenta
\begin{align}
 A_{11,k}^{(m)}={}&\frac{(\pi_1^\xi)^2}{2a^2\epsilon}+\frac{(\pi_1^\psi)^2}{a^2}
 +2a^2\epsilon k^2\xi_1^2+a^2(k^2+a^2\mu^2)\psi_1^2
 +\frac{2\mathcal H\lambda}{\sqrt{2\epsilon}}\pi_1^\xi\psi_1,\label{eq:Am11}\\
 A_{12,k}^{(m)}={}&\frac{\pi_1^\xi\pi_2^\xi}{2a^2\epsilon}
 +\frac{\pi_1^\psi\pi_2^\psi}{a^2}
 +2a^2\epsilon k^2\xi_1\xi_2+a^2(k^2+a^2\mu^2)\psi_1\psi_2
 \nonumber\\&+\frac{\mathcal H\lambda}{\sqrt{2\epsilon}}
 (\pi_1^\xi\psi_2+\psi_1\pi_2^\xi),\label{eq:Am12}\\
 A_{21,k}^{(m)}={}&\frac{\pi_2^\xi\pi_1^\xi}{2a^2\epsilon}
 +\frac{\pi_2^\psi\pi_1^\psi}{a^2}
 +2a^2\epsilon k^2\xi_2\xi_1+a^2(k^2+a^2\mu^2)\psi_2\psi_1
 \nonumber\\&+\frac{\mathcal H\lambda}{\sqrt{2\epsilon}}
 (\pi_2^\xi\psi_1+\psi_2\pi_1^\xi)=A_{12}^{(m)},\label{eq:Am21}\\
 A_{22,k}^{(m)}={}&\frac{(\pi_2^\xi)^2}{2a^2\epsilon}+\frac{(\pi_2^\psi)^2}{a^2}
 +2a^2\epsilon k^2\xi_2^2+a^2(k^2+a^2\mu^2)\psi_2^2
 +\frac{2\mathcal H\lambda}{\sqrt{2\epsilon}}\pi_2^\xi\psi_2.
 \label{eq:Am22}
\end{align}
The canonical momenta used here are fully specified by $\pi_i^\xi=2a^2\epsilon[\xi_i'-(\mathcal H\lambda/\sqrt{2\epsilon})\psi_i]$ and $\pi_i^\psi=a^2\psi_i'$. The four number-conserving coefficients are
\begin{align}
 B_{11,k}^{(m)}={}&\frac{|\pi_1^\xi|^2}{2a^2\epsilon}+\frac{|\pi_1^\psi|^2}{a^2}
 +2a^2\epsilon k^2|\xi_1|^2+a^2(k^2+a^2\mu^2)|\psi_1|^2
 +\frac{\mathcal H\lambda}{\sqrt{2\epsilon}}(\pi_1^{\xi*}\psi_1+\psi_1^*\pi_1^\xi),\label{eq:Bm11}\\
 B_{12,k}^{(m)}={}&\frac{\pi_1^{\xi*}\pi_2^\xi}{2a^2\epsilon}
 +\frac{\pi_1^{\psi*}\pi_2^\psi}{a^2}
 +2a^2\epsilon k^2\xi_1^*\xi_2+a^2(k^2+a^2\mu^2)\psi_1^*\psi_2
 \nonumber\\&+\frac{\mathcal H\lambda}{\sqrt{2\epsilon}}
 (\pi_1^{\xi*}\psi_2+\psi_1^*\pi_2^\xi),\label{eq:Bm12}\\
 B_{21,k}^{(m)}={}&\frac{\pi_2^{\xi*}\pi_1^\xi}{2a^2\epsilon}
 +\frac{\pi_2^{\psi*}\pi_1^\psi}{a^2}
 +2a^2\epsilon k^2\xi_2^*\xi_1+a^2(k^2+a^2\mu^2)\psi_2^*\psi_1
 \nonumber\\&+\frac{\mathcal H\lambda}{\sqrt{2\epsilon}}
 (\pi_2^{\xi*}\psi_1+\psi_2^*\pi_1^\xi)=(B_{12}^{(m)})^*,\label{eq:Bm21}\\
 B_{22,k}^{(m)}={}&\frac{|\pi_2^\xi|^2}{2a^2\epsilon}+\frac{|\pi_2^\psi|^2}{a^2}
 +2a^2\epsilon k^2|\xi_2|^2+a^2(k^2+a^2\mu^2)|\psi_2|^2
 +\frac{\mathcal H\lambda}{\sqrt{2\epsilon}}(\pi_2^{\xi*}\psi_2+\psi_2^*\pi_2^\xi).
 \label{eq:Bm22}
\end{align}
The mode functions obey the fully coupled equations
\begin{align}
 &~~~~~~\xi_i'=\frac{\pi_i^\xi}{2a^2\epsilon}+\frac{\mathcal H\lambda}{\sqrt{2\epsilon}}\psi_i,\qquad \psi_i'=\frac{\pi_i^\psi}{a^2}, \qquad i=1,2,\nonumber\\
& (\pi_i^\xi)'=-2a^2\epsilon k^2\xi_i,\qquad
 (\pi_i^\psi)'=-a^2(k^2+a^2\mu^2)\psi_i
 -\frac{\mathcal H\lambda}{\sqrt{2\epsilon}}\pi_i^\xi.
 \label{eq:modeODE}
\end{align}
The superscript $(m)$ specifies the time-dependent mode-function frame. In that frame, these expressions represent the Hamiltonian written through initial, fixed operators. They are not coefficients to substitute into Eqs.~\eqref{eq:xODE}--\eqref{eq:zODE}, which use the fixed canonical vacuum of Eq.~\eqref{eq:realH}. Mixing the two frames without their time-dependent canonical connection would change the state equation.

\section{Occupation amplitudes and extraction of four-mode squeezed state}
\label{app:state}

\subsection{Exact four-mode Fock expansion}
Let $|N_{1,+},N_{1,-},N_{2,+},N_{2,-}\rangle$ denote the normalized four-mode number state used in Ref.~\cite{Colas:2021llj}. Expanding each of the four terms in Eq.~\eqref{eq:fourstate} independently gives an explicit, untruncated sum
\begin{align}
 |\Psi_k\rangle={}&\mathcal N_k
 \sum_{n_{ij}=0}^{\infty}
 \frac{Z_{11,k}^{n_{11}}Z_{12,k}^{n_{12}+n_{21}}Z_{22,k}^{n_{22}}}
 {n_{11}!n_{12}!n_{21}!n_{22}!}\nonumber\\
 &\quad\times\sqrt{(n_{11}+n_{12})!(n_{11}+n_{21})!
 (n_{21}+n_{22})!(n_{12}+n_{22})!}\nonumber\\
 &\quad\times|n_{11}+n_{12},\,n_{11}+n_{21},\,
 n_{21}+n_{22},\,n_{12}+n_{22}\rangle.
 \label{eq:fockseries}
\end{align}
This realizes the same four-mode sectors and transfer processes as the occupation-number expansion in Eqs.~(5.6)--(5.7) of Ref.~\cite{Colas:2021llj}. Distinct quadruples $(n_{11},n_{12},n_{21},n_{22})$ can yield the same Fock vector. For a prescribed $(N_{1,+},N_{1,-},N_{2,+},N_{2,-})$, all solutions of the four integer constraints appearing in Eq.~\eqref{eq:fockseries} must be summed before a physical Fock amplitude is read off. In particular
\begin{align}
 \langle0,0,0,0|\Psi_k\rangle&=\mathcal N_k,&
 \langle1,1,0,0|\Psi_k\rangle&=\mathcal N_k Z_{11,k},\nonumber\\
 \langle1,0,0,1|\Psi_k\rangle&=\mathcal N_k Z_{12,k},&
 \langle0,1,1,0|\Psi_k\rangle&=\mathcal N_k Z_{12,k},\nonumber\\
 \langle0,0,1,1|\Psi_k\rangle&=\mathcal N_k Z_{22,k},&
 \langle1,1,1,1|\Psi_k\rangle&=\mathcal N_k(Z_{11,k}Z_{22,k}+Z_{12,k}^2).
 \label{eq:lowFock}
\end{align}
Hence the three complex amplitudes can be extracted from the original four-mode wavefunction without a mode transformation
\begin{equation}
 Z_{11,k}=\frac{\langle1,1,0,0|\Psi_k\rangle}{\langle0,0,0,0|\Psi_k\rangle},\quad
 Z_{12,k}=\frac{\langle1,0,0,1|\Psi_k\rangle}{\langle0,0,0,0|\Psi_k\rangle}
 =\frac{\langle0,1,1,0|\Psi_k\rangle}{\langle0,0,0,0|\Psi_k\rangle},\quad
 Z_{22,k}=\frac{\langle0,0,1,1|\Psi_k\rangle}{\langle0,0,0,0|\Psi_k\rangle}.
 \label{eq:extractFock}
\end{equation}
The coefficients $c_k(n,m,s,t)$ of Eq.~(5.7) in Ref.~\cite{Colas:2021llj} may also be used directly. Their state indices in Eq.~(5.6) are $(n+s,n+t,m-s,m-t)$. For any occupation vector satisfying $N_{1,+}+N_{2,+}=N_{1,-}+N_{2,-}=L$, its actual amplitude is the following sum, with no Fock-space truncation
\begin{equation}
 \langle N_{1,+},N_{1,-},N_{2,+},N_{2,-}|\Psi_k\rangle
 =\sum_{n=0}^{L}c_k\bigl(n,L-n,N_{1,+}-n,N_{1,-}-n\bigr).
 \label{eq:originalsum}
\end{equation}
Consequently, the three ratios in Eq.~\eqref{eq:extractFock} can be computed using the original paper's explicit coefficients as
\begin{align}
 Z_{11,k}&=\frac{c_k(0,1,1,1)+c_k(1,0,0,0)}{c_k(0,0,0,0)},\nonumber\\
 Z_{12,k}&=\frac{c_k(0,1,1,0)+c_k(1,0,0,-1)}{c_k(0,0,0,0)},\nonumber\\
 Z_{22,k}&=\frac{c_k(0,1,0,0)+c_k(1,0,-1,-1)}{c_k(0,0,0,0)}.
 \label{eq:originalratios}
\end{align}
These formulas expose the multiple contributions that would be missed by taking a ratio of single $c_k$ terms. The four-particle example in Eq.~\eqref{eq:lowFock} provides a second consistency check. The phases of the $c_k$ coefficients include an arbitrary overall vacuum phase, which cancels from all three ratios.

For completeness, substituting the original coefficients into Eq.~\eqref{eq:originalratios} gives their explicit relation to the original rotation and squeezing parameters. Use the paper's $\theta_{3k},\theta_{4k},\theta_{5k},\theta_{6k}$ and define the original ordered-product parameters, with their full formulas displayed
\begin{align}
& ~~~~~~~~~~~~~~~~~~~~~~~\vartheta_k^{\rm(orig)}=\sqrt{\theta_{4k}^2+\theta_{5k}^2+\theta_{6k}^2},
 ~~~~~~~\tau_k^{\rm(orig)}=-\theta_{6k}+i\theta_{5k},\nonumber\\
& p_{z,k}=-2\log\!\left[\cos\vartheta_k^{\rm(orig)}
 -i\frac{\theta_{4k}}{\vartheta_k^{\rm(orig)}}\sin\vartheta_k^{\rm(orig)}\right],
 p_{-,k}=\frac{-[\tau_k^{\rm(orig)}]^*\sin\vartheta_k^{\rm(orig)}}
 {\vartheta_k^{\rm(orig)}\cos\vartheta_k^{\rm(orig)}
 -i\theta_{4k}\sin\vartheta_k^{\rm(orig)}},\nonumber\\
&~~~~~~~~~~~~~~~~~~~~~~~~~~~~~ p_{+,k}=\frac{\tau_k^{\rm(orig)}\sin\vartheta_k^{\rm(orig)}}
 {\vartheta_k^{\rm(orig)}\cos\vartheta_k^{\rm(orig)}
 -i\theta_{4k}\sin\vartheta_k^{\rm(orig)}}.
 \label{eq:originalp}
\end{align}
The explicit amplitudes obtained by summing the two paths in each numerator are
\begin{align}
& Z_{11,k}={}-e^{-2i\theta_{3k}}e^{-p_{z,k}}
 \left[\tanh r_{1k}^{\rm(orig)}-p_{-,k}^2\tanh r_{2k}^{\rm(orig)}\right],\label{eq:xoriginal}\\
 Z_{12,k}={}&i e^{-2i\theta_{3k}}\Bigl[p_{-,k}\tanh r_{2k}^{\rm(orig)}
 \left(1+p_{-,k}p_{+,k}e^{-p_{z,k}}\right)
 -e^{-p_{z,k}}p_{+,k}\tanh r_{1k}^{\rm(orig)}\Bigr],\label{eq:yoriginal}\\
 Z_{22,k}={}&e^{-2i\theta_{3k}}\Bigl[e^{-p_{z,k}}p_{+,k}^{\,2}
 \tanh r_{1k}^{\rm(orig)}
 -e^{p_{z,k}}\left(1+p_{-,k}p_{+,k}e^{-p_{z,k}}\right)^2
 \tanh r_{2k}^{\rm(orig)}\Bigr].
 \label{eq:zoriginal}
\end{align}
The common vacuum phase associated with the paper's right rotation cancels. The original decomposition permits signed squeezing coordinates, whereas Eqs.~\eqref{eq:r1}--\eqref{eq:r2} select nonnegative strengths and absorb their signs into passive phases. The smooth zero-angle limit of Eq.~\eqref{eq:originalp} is understood when $\vartheta_k^{\rm(orig)}=0$. In the inflationary model, it is simpler to integrate Eqs.~\eqref{eq:xODE}--\eqref{eq:zODE} and then evaluate the fully explicit strengths. Eqs.~\eqref{eq:xoriginal}--\eqref{eq:zoriginal} provide a direct bridge to the original paper's four-mode parametrization.

\subsection{Direct solution for both squeezing strengths and passive phases}
The four-mode squeezed state is normalizable if and only if the two eigenvalues of the positive matrix
\begin{equation}
 \begin{pmatrix}
 |Z_{11,k}|^2+|Z_{12,k}|^2&Z_{11,k}Z_{12,k}^*+Z_{12,k}Z_{22,k}^*\\
 Z_{11,k}^*Z_{12,k}+Z_{12,k}^*Z_{22,k}&|Z_{22,k}|^2+|Z_{12,k}|^2
 \end{pmatrix}
 \label{eq:Hamp}
\end{equation}
lie below one. The normalized passive field combinations are obtained by diagonalizing Eq.~\eqref{eq:Hamp}. Its characteristic equation is explicitly
\begin{align}
 0&=u^2-(|Z_{11,k}|^2+|Z_{22,k}|^2+2|Z_{12,k}|^2)u\nonumber\\
 &+(|Z_{11,k}|^2+|Z_{12,k}|^2)(|Z_{22,k}|^2+|Z_{12,k}|^2)\nonumber\\
 &-|Z_{11,k}Z_{12,k}^*+Z_{12,k}Z_{22,k}^*|^2,\nonumber\\
 &=u^2-(|Z_{11,k}|^2+|Z_{22,k}|^2+2|Z_{12,k}|^2)u\\&+|Z_{11,k}Z_{22,k}-Z_{12,k}^2|^2.
 \label{eq:charpoly}
\end{align}
Solving this quadratic produces the squared arguments of the inverse hyperbolic tangents in Eqs.~\eqref{eq:r1} and~\eqref{eq:r2}. The two-field rotation and the two phase-bearing squeezers of Eq.~\eqref{eq:stateBM} give the three expanded identities \eqref{eq:Z11physical}--\eqref{eq:Z22physical}. Because the rotation in Eq.~\eqref{eq:Zsix} is unitary, their product with the complex-conjugate transpose gives
\begin{equation}
 Z_kZ_k^\dagger=U_k
 \begin{pmatrix}\tanh^2r_{1k}&0\\0&\tanh^2r_{2k}\end{pmatrix}U_k^\dagger.
 \label{eq:Takagiproof}
\end{equation}
Thus the eigenvalues of Eq.~\eqref{eq:Hamp} are precisely $\tanh^2r_{1k}$ and $\tanh^2r_{2k}$, independent of how the original field labels are mixed.

The remaining state parameters are also explicit. For unequal eigenvalues set $h_{11}=|Z_{11,k}|^2+|Z_{12,k}|^2$, $h_{22}=|Z_{22,k}|^2+|Z_{12,k}|^2$, $h_{12}=Z_{11,k}Z_{12,k}^*+Z_{12,k}Z_{22,k}^*$ and choose
\begin{equation}
 \chi_k=\arg h_{12},\quad
 \vartheta_k=\frac12\operatorname{atan2}(2|h_{12}|,h_{11}-h_{22}),\quad
 U_{0,k}=\begin{pmatrix}
 \cos\vartheta_k&-e^{i\chi_k}\sin\vartheta_k\\
 e^{-i\chi_k}\sin\vartheta_k&\cos\vartheta_k
 \end{pmatrix}.
 \label{eq:mixing}
\end{equation}
The two complex diagonal amplitudes in this rotated basis are
\begin{align}
 w_{1k}&=Z_{11,k}\cos^2\vartheta_k
 +2Z_{12,k}e^{i\chi_k}\sin\vartheta_k\cos\vartheta_k
 +Z_{22,k}e^{2i\chi_k}\sin^2\vartheta_k,\label{eq:w1}\\
 w_{2k}&=Z_{11,k}e^{-2i\chi_k}\sin^2\vartheta_k
 -2Z_{12,k}e^{-i\chi_k}\sin\vartheta_k\cos\vartheta_k
 +Z_{22,k}\cos^2\vartheta_k.
 \label{eq:w2}
\end{align}
In the angle convention of Ref.~\cite{Li:2026jxx}, the phase of the pair coefficient is $-e^{2i\phi_{ik}}$. Then, we could obtain
\begin{equation}
 \begin{aligned}
 |w_{1k}|&=\tanh r_{1k},&
 \phi_{1k}&=\tfrac12\arg(-w_{1k})\pmod\pi,\\
 |w_{2k}|&=\tanh r_{2k},&
 \phi_{2k}&=\tfrac12\arg(-w_{2k})\pmod\pi.
 \end{aligned}
 \label{eq:phaseextract}
\end{equation}
The phase choice in Eq.~\eqref{eq:mixing} fixes the otherwise movable phase convention of the passive transformation. When $h_{12}=0$, choose the coordinate eigenvectors directly, at a degenerate eigenvalue the mixing basis is nonunique. Neither situation changes the strengths or complexity. At $r_{ik}=0$, its associated phase is arbitrary.

For a direct differential comparison with the single-channel $r_k',\phi_k'$ equations of Ref.~\cite{Li:2026jxx}, differentiating the preceding identities gives an explicit six-real-parameter evolution after Eqs.~\eqref{eq:xODE}--\eqref{eq:zODE} have supplied the three complex derivatives. The required entries are
\begin{align}
& ~~~~~~~~~~~~~~h_{11}'=2\text{Re} \!\left(Z_{11,k}^*Z_{11,k}'+Z_{12,k}^*Z_{12,k}'\right),\nonumber\\
&~~~~~~~~~~~~~~ h_{22}'=2\text{Re}\!\left(Z_{22,k}^*Z_{22,k}'+Z_{12,k}^*Z_{12,k}'\right),\nonumber\\
& h_{12}'=Z_{11,k}'Z_{12,k}^*+Z_{11,k}(Z_{12,k}')^*
 +Z_{12,k}'Z_{22,k}^*+Z_{12,k}(Z_{22,k}')^*.
 \label{eq:hderivatives}
\end{align}
For nondegenerate positive squeezing strengths, set $\Delta_k=\sqrt{(h_{11}-h_{22})^2+4|h_{12}|^2}$. The strength equations are
\begin{align}
&~~~~~~~ r_{1k}'={}\frac{h_{11}'+h_{22}'+\Delta_k'}
 {4\tanh r_{1k}\,(1-\tanh^2r_{1k})},\nonumber\\
& ~~~~~~~r_{2k}'={}\frac{h_{11}'+h_{22}'-\Delta_k'}
 {4\tanh r_{2k}\,(1-\tanh^2r_{2k})},\nonumber\\
& \Delta_k'={}\frac{(h_{11}-h_{22})(h_{11}'-h_{22}')
 +4\text{Re}(h_{12}^*h_{12}')}{\Delta_k}.
 \label{eq:rderivatives}
\end{align}
The mixing and squeezing angle derivatives are
\begin{align}
& ~~~~~~~~~~~~~~~~~~~~~~~\chi_k'=\text{Im}\!\left(\frac{h_{12}'}{h_{12}}\right),\nonumber\\
& \vartheta_k'=\frac{(h_{11}-h_{22})\text{Re}(h_{12}^*h_{12}')/|h_{12}|
 -|h_{12}|(h_{11}'-h_{22}')}
 {(h_{11}-h_{22})^2+4|h_{12}|^2},\nonumber\\
&~~~~~~ \phi_{1k}'=\frac12\text{Im}\!\left(\frac{w_{1k}'}{w_{1k}}\right),\qquad
 \phi_{2k}'=\frac12\text{Im}\!\left(\frac{w_{2k}'}{w_{2k}}\right),
 \label{eq:anglederivatives}
\end{align}
where $w_{1k}'$ and $w_{2k}'$ mean the total conformal-time derivatives of the explicit expressions \eqref{eq:w1}--\eqref{eq:w2}, using Eqs.~\eqref{eq:xODE}--\eqref{eq:zODE}, the derivatives of \(h_{ij}\) from Eq. \eqref{eq:hderivatives}, and the mixing-parameter derivatives given above. These phase-coordinate equations become singular at zero squeezing or degenerate strengths even though the original three complex equations remain regular. The latter are the preferred initial-value system. Together they determine the physical $r_{ik}$ and $\phi_{ik}$ without discarding either cross-field Hamiltonian coefficient.

\subsection{Component evolution from the Schr\"odinger equation}
The four identities in Eq.~\eqref{eq:annihilation} reduce every annihilation term in Eq.~\eqref{eq:pairH} to a vacuum term and four creation channels. The coefficient of $b_{1,+}^\dagger b_{1,-}^\dagger$ on the left is $i\mathcal N_k Z_{11,k}'$, that of either cross pair is $i\mathcal N_k Z_{12,k}'$, and that of $b_{2,+}^\dagger b_{2,-}^\dagger$ is $i\mathcal N_k Z_{22,k}'$. Matching each channel separately gives Eqs.~\eqref{eq:xODE}--\eqref{eq:zODE}. Matching the vacuum channel gives Eq.~\eqref{eq:NODE}. No off-diagonal transfer or cross-pair contribution is omitted. In a deeply subhorizon initial state the finite-time vacuum condition $Z_{11,k}(\tau_i)=Z_{12,k}(\tau_i)=Z_{22,k}(\tau_i)=0$ approximates the asymptotic Bunch--Davies condition. A different initial Gaussian state requires its own initial amplitudes.

\acknowledgments
T.L. acknowledges support from the Guizhou Provincial Natural Science Foundation (Grant No. 2025YJSKYJJ178) and the Natural Science Foundation of Guizhou Minzu University (Grant No. GZMUBS7). L.H. acknowledges support from the National Natural Science Foundation of China (Grant No. 12665009) and the Hunan Provincial Department of Education Project (Grant No. 25B0480). H.B. acknowledges support from the National Natural Science Foundation of China (Grant No. 12265010) and the Guizhou Provincial Department of Science and Technology (Grant Nos. MS[2025]219 and CXTD[2025]030).

\bibliography{refs.bib}

\end{document}